\documentclass[a4paper,11pt]{article}
\pdfoutput=1 % if your are submitting a pdflatex (i.e. if you have
\usepackage[utf8]{inputenc}
\usepackage{jinstpub} % for details on the use of the package, please
\usepackage{hyperref}
\usepackage{epstopdf}
\usepackage{xcolor}
\usepackage{graphicx}
\usepackage{subfigure}
\usepackage{amsmath}
\usepackage{breqn}
\usepackage{comment}

\newcommand{\executeiffilenewer}[3]{%
 \ifnum\pdfstrcmp{\pdffilemoddate{#1}}%
 {\pdffilemoddate{#2}}>0%
 {\immediate\write18{#3}}\fi%
}

\newcommand{\includesvg}[1]{%
 \executeiffilenewer{#1.svg}{#1.pdf}%
 {inkscape -z -D --file=#1.svg %
 --export-pdf=#1.pdf --export-latex}%
}

\title{\boldmath A High Precision Pulse Generation and Stabilization System for Bolometric Experiments}

\author[a,b,1]{P. Carniti \note{Corresponding author.},}
\author[a,b]{L. Cassina,}
\author[a,b]{A. Giachero,}
\author[a,b,1]{C. Gotti,}
\author[a,b]{G. Pessina}

\affiliation[a]{INFN, Sezione di Milano Bicocca,
Piazza della Scienza 3, 20126, Milano, Italy}
\affiliation[b]{ Dipartimento di Fisica G. Occhialini, Universit\`a degli Studi di Milano Bicocca,
Piazza della Scienza 3, 20126, Milano, Italy}

\emailAdd{paolo.carniti@mib.infn.it, claudio.gotti@mib.infn.it}

\abstract{Bolometric experiments searching for rare events usually require an extremely low radioactive background to prevent spurious signals from mimicking those of interest, spoiling the sensitivity of the apparatus. In such contexts, radioactive sources cannot be used to produce a known signal to calibrate the measured energy spectrum during data taking. In this paper we present an instrument designed to generate ultra-stable and very precise calibrating pulse, which can be used to stabilize the response of bolometers during data taking. The instrument is characterized by the presence of multi-outputs, a completely programmable pulse width and amplitude and a dedicated daisy-chained optical trigger line. It can be fully controlled and monitored remotely via CAN bus protocol. An energy resolution of the order of 20 eV FWHM at 1 MeV (2 eV FWHM at 10 keV) and a thermal stability of the order of 0.1 ppm/$^\circ$C have been achieved. The device can also provide an adjustable power to compensate the low frequency thermal fluctuations that typically occur in cryogenic experiments.}

\keywords{Analogue electronic circuits, Detector alignment and calibration methods (lasers, sources, particle-beams), Double-beta decay detectors}

\begin{document}
\maketitle
\flushbottom

\section{Introduction}\label{sec:intro}
The search for rare events is of great importance in modern particle physics, as it allows to thoroughly test Standard Model predictions. In particular, a powerful tool to study neutrino physics and dark matter consists of looking for very rare spontaneous nuclear decays or nuclear scattering due to interaction with dark matter candidates.% a Weak Interacting Massive Particle (WIMP).
Such class of experiments, which usually need long live time (several years) in order to collect adequate statistics, typically requires an extremely low radioactive background and an excellent energy resolution. Among the various experimental approaches, bolometric detectors stand out as they ensure very good intrinsic energy resolution, no dead regions, a wide choice of materials and the potential to be arranged in large arrays of detectors. Bolometers are crystal calorimeters operating at cryogenic temperature in which the energy released by an interacting particle is measured from the consequent temperature increase.

%One of the most important features that must be ensured by the electronic equipment is stability over the whole live time.
%As the gain of a bolometer strongly depends on its operating temperature, its response must be stabilized against the low frequency thermal instabilities which unavoidably affect the cryogenic system.  %which can spoil the energy conversion gain and its resolution. 
As the gain of a bolometer strongly depends on its operating temperature, its working temperature must be stabilized against low frequency fluctuations of the cryogenic system, and any residual temperature drift must be precisely tracked and compensated for offline.
An absolute calibration can be performed periodically by using a known $\gamma$-ray calibrating source, but not during data taking, due to the requirements on low radioactive background.
During data taking, particle interactions can be mimicked by delivering triggered voltage pulses across resistive heaters glued to the bolometers, so that thermal power can be produced by Joule effect \cite{ArtCalib}.
The calibrating signal is a square pulse having a width ($T$) much shorter than the detector response time, tipically $\sim100$~ms for large bolometers. % given by the product between the crystal thermal capacitance (few nJ$/^\circ$K) and the thermal conductivity towards the cooling reference. 
With this method, the pulse amplitude and width can be tuned to stabilize each detector over the full dynamic range.
The pulses are triggered, allowing to generate them with a selectable rate and easily tag them.
While the power consumed on the heating resistor is controlled by the pulse parameters, the energy injected to the bolometers is proportional to the thermal coupling between the heating resistor and the crystal ($\eta$).
This constant is not known a priori and changes from crystal to crystal, therefore the pulses generated by Joule effect cannot provide an absolute calibration.
The efficacy of such method has already been demonstrated in the CUORICINO \cite{CUORICINO} experiment, where a high-stability pulse generator has been used \cite{PUL1,PUL2}. 

%The solution proposed exploits the benefits of both approaches. The radioactive source is used to initially calibrate the system, while the heating source is used to stabilize it during the data taking. In particular, the heater pulses are used to monitor the bolometer gain as a function of its operating temperature, measured from the detector baseline\cite{ArtCalib}. With this procedure, the amplitude of each pulse is rescaled as a function of its baseline, allowing to compensate for the temperature thermal drift  and to maintain the energy resolution at its optimum value for the entire data taking. The efficacy of such method has already been demonstrated in the CUORICINO\cite{CUORICINO} experiment, where a high-stability pulse generator has been used\cite{PUL1,PUL2}. 

In this paper we present a novel, multi-channel, ultra-stable and ultra-precise electronic system designed for the stabilization of bolometer response. This new equipment is currently used in the CUORE \cite{CUORE1,CUORE2}, LUCIFER/CUPID-0 \cite{LUCIFER,CUPID} and COSINUS \cite{COSINUS} experiments, all located underground at Laboratori Nazionali del Gran Sasso, Italy. In section \ref{sec:PulserDesign} the description of the circuit is provided. In order to achieve high energy resolution and stability, particular care was addressed to obtain low noise and minimize drift against temperature. The thermal calibration of the circuit and its stability performance are shown in section \ref{sec:Stab}. The results on pulse accuracy, jitter and electrical noise are presented in section \ref{sec:Accuracy}. The device can also been used to optimize the cryogenic temperature of operation of the bolometers and stabilize it against slow drifts: in section \ref{sec:PID} a detailed description of such architecture and the results achieved in the experimental context are shown.

\section{Circuit description}\label{sec:PulserDesign}
%

%\begin{figure}[h!]
%	\centering
%	\includegraphics[width=1\textwidth]{Immagini/LayoutPulser.pdf}
%	\caption{A picture of the layout of the CUORE pulser board. The main functional components are pointed out.}
%	\label{fig:Layout}
%\end{figure}

The pulser board was specifically designed to provide calibrating pulses with outstanding stability, extremely low noise, high timing precision and high reliability.
 Each board has four output channels.
 The design includes two input/output optical converters to receive an optional external trigger and share it in a daisy chain configuration.

The device consists of a four layer Printed Circuit Board hosting a Cortex M3 32-bit microcontroller (LPC1768 from NXP) managing all the on-board peripherals. The microcontroller communicates with the remote control system through a 100 kbit/s CAN bus. In such industrial protocol each command is proceeded by a 11/29 bit code which uniquely identifies the desired board, sharing the bus with all the others. Thus, a single link is enough to connect all the needed boards, minimizing the cabling budget. High reliability is guaranteed by the fact that each node on the bus actively contributes to detect communication errors and correct them. %In the experimental case, an optional additional board mediates the CAN communication in order to optically decoulpe the pulser board to the noisy remote control system.

\subsection{Pulse generation stage}\label{sec:OutputBlock}

\begin{figure}[t]
	\centering
	\includegraphics[width=1\textwidth]{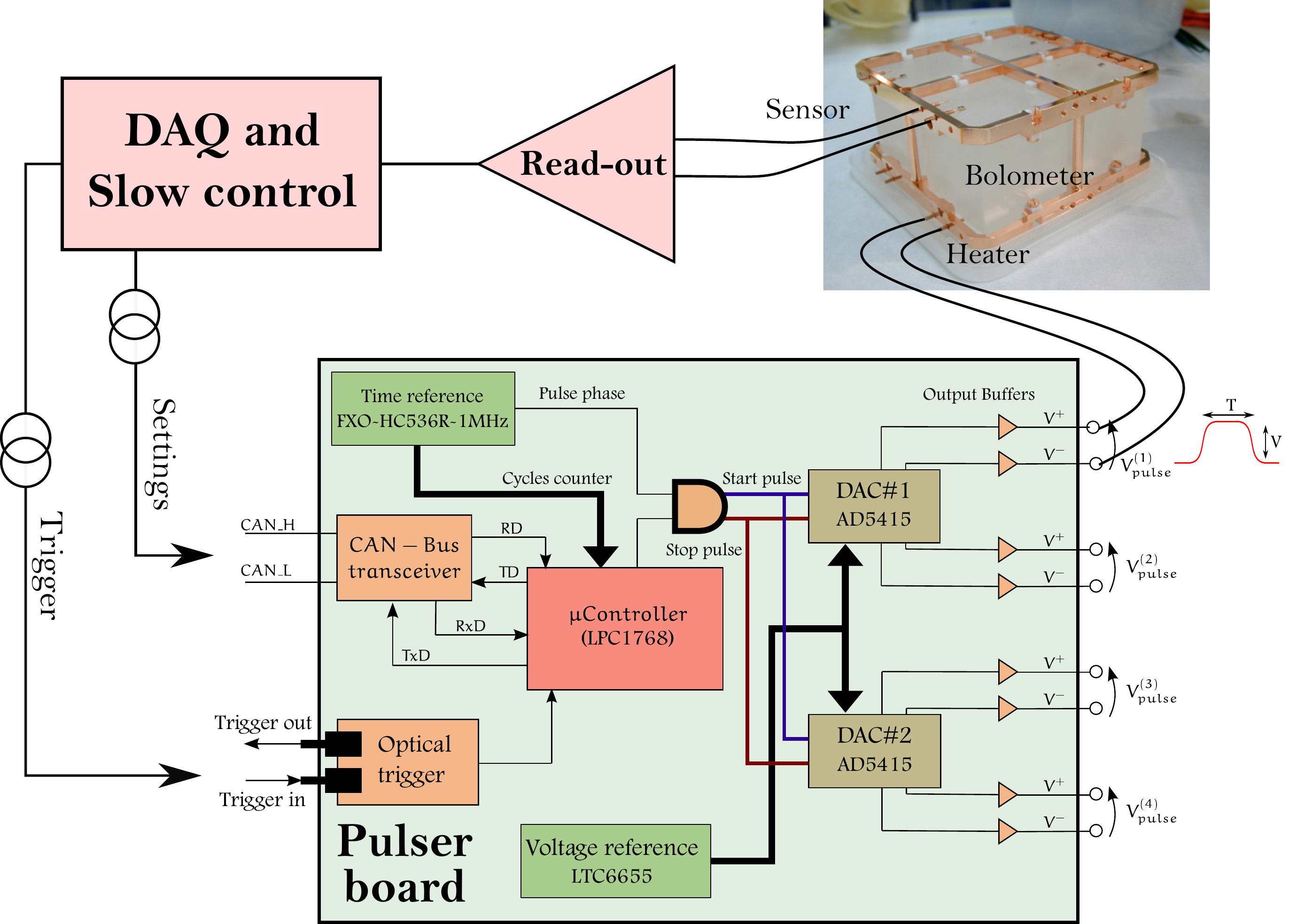}
	\caption{Simplified scheme to calibrate and stabilize the bolometer.}
	\label{fig:SchemaABlocchi}
\end{figure}

Figure \ref{fig:SchemaABlocchi} shows the main blocks composing the pulse generation chain for a typical application such as the CUORE experiment. The high precision differential square pulse is routed into the cryostat with twisted pairs and reaches the heater resistors glued to the bolometers. The energy injected on the bolometer, expressed in $eV$, is given by

\begin{equation}\label{eq:Joule}
E[eV]=\eta\frac{V_0^2 T}{e R \varepsilon^2}
\end{equation}
where $0 \leq \eta \leq 1$ is the thermal coupling between the crystal and the heater, $R$ is the heater resitance, $V_0$ and $T$ are the amplitude and width of the square pulse, $\varepsilon$ the attenuation factor due to the output voltage divider (as explained below) and $e$ is the electron charge.
%The energy released to the bolometer is converted into a voltage signal by the thermal sensor (such as a Neutron Transmutation Doping thermistor, in the case of CUORE and CUPID-0 experiments), amplified by the front-end electronics and digitized by the DAQ system. 

In order to easily tag the calibrating pulse and prevent synchronization issues, it is convenient that the DAQ system triggers the pulse delivery after the pulse parameters are set by the slow control system. % instead of the slow control system, responsible to set the pulse parameters. 
In particular, the DAQ provides a digital TTL transition which is converted into an optical pulse and driven to the pulser board front panel, hosting an optical receiver triggering the final pulse generation. The trigger is provided by an optical signal to completely decouple the ground of the DAQ system from the pulser board, avoiding ground loops. The optical trigger can also be daisy chained among several boards, so that a unique optical connection from the DAQ system to the first pulser is enough to provide the trigger to the whole system.
%This is particularly useful as the pulser boards are usually located far from the DAQ system, close to the cryostat and possibly inside a Faraday cage.  \\

Figure \ref{fig:SchemaABlocchi} also shows the main blocks used to generate the calibrating pulse. In order to obtain precise pulse transitions, the pulser board is equipped with a FXO-HC536R oscillator from FOX Electronics, able to provide a 1 MHz reference clock with a declared phase stability of $\pm$25 ppm over its whole temperature range and a jitter better than a few tens of ps RMS. This time reference is connected at the input of the LPC1768 microcontroller, which can enable or disable the clock reference and count its cycles. The first order estimation of the pulse width is obtained simply by converting the desired duration in terms of equivalent clock cycles (i.e. as an integer multiple of the 1 $\mu$s time unit). However, the timing performance of the microcontroller is not good enough to directly trigger the pulse transitions. Instead, the voltage generator is triggered by a fast flip-flop (the NC7SV74 from Fairchild Semiconductor) connected to both the microcontroller and the reference oscillator. The microcontroller acts as a gate: when the gate is closed, the flip-flop maintains the output voltage generator in reset mode. The gate is opened as soon as the microcontroller clock cycles counter matches the target value (i.e. the pulse coarsely lasts as desired). When the gate is open, the flip-flop triggers the pulse transitions at the next rising edge of the reference clock. Using such architecture, the timing performance only depends on the precise clock reference and the fast flip-flop performance. In particular, the phase stability of the generated pulse coincides with that of the reference clock irrespectively of the pulse width and the microcontroller response. The outcomes of the timing performance characterization in term of temperature stability and random fluctuations will be shown in sections \ref{sec:Stab} and \ref{sec:Accuracy} respectively. The board is equipped with two other clock sources which serve as reference for the microcontroller operation and for the on-board analog to digital coverter (ADC). In order to prevent the fast transitions of the clocks from inducing high frequency disturbances, all these components are enclosed inside a grounded metallic box acting as a local Faraday cage.\\ 
%Figure \ref{fig:pulse} shows a typical square pulse generated at the output of the pulser board.\\

\begin{figure}[t] 
\centering
  \def\svgwidth{0.9\columnwidth}
  \graphicspath{{Immagini/}}
  \includesvg{Immagini/SchemaPulser}
  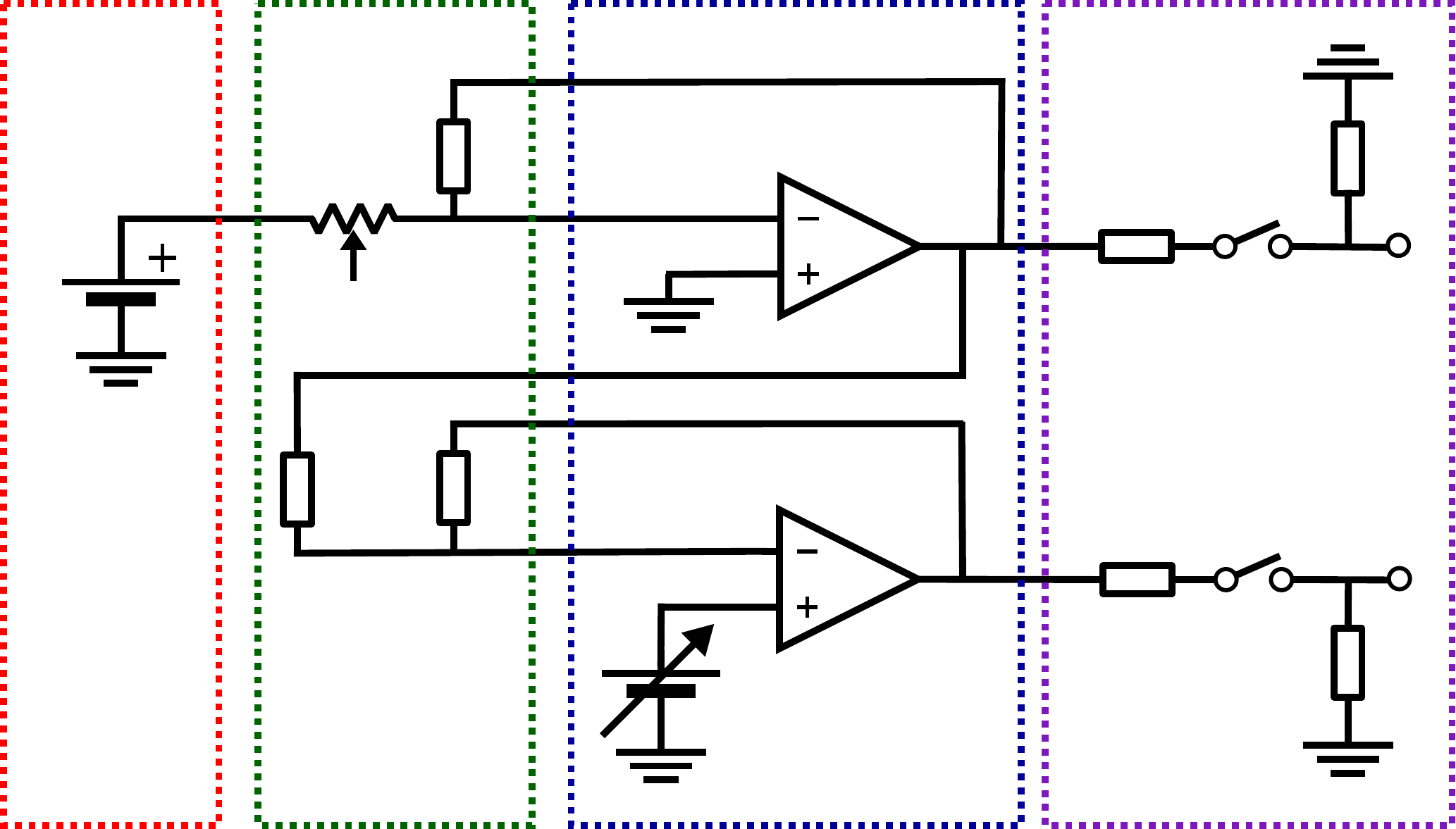%
  \caption{Simplified schematic of a pulser board channel}
  \label{fig:Scheme} 
\end{figure}

The output voltage generator system is based on the very precise and low noise voltage reference LTC6655 from Linear Technology. This component generates a 5 V voltage reference ensuring a high thermal stability (few ppm$/^\circ$C) and a negligible sensitivity to the humidity level if packaged in the LS8 hermetic case. In order to reduce its noise contribution by $30\%$, two components are connected in parallel and averaged, and the resulting voltage is used to feed the AD5415, a 12-bit, dual channel digital to analog converter (DAC) from Analog Devices. Coupled with the low noise and high precision OPA140 operational amplifier, from Texas Instruments, the DAC can be schematized as a settable resistor, ranging from 10 k$\Omega$ to an open circuit, as shown in figure \ref{fig:Scheme}. 

Since $R_{D1}$ amounts to 10 k$\Omega$ and the OPA140 is used in inverting configuration, this architecture delivers a negative voltage level ranging from ground to $-V_{REF}=-5$ V. This signal is also driven at the input of an inverting unity gain amplifier, composed of a second OPA140 operational amplifier and the $R_{D2}=R_{D3}=20$ k$\Omega$ resistors embedded in the DAC. As a result, a differential signal up to 10 V can be supplied between the outputs. Capacitors $C_1$ and $C_3$ limit the bandwidth of the device to $1.3$~MHz, to increase the output buffer phase margin and avoid ringing. Note that the positive terminal of the second inverting amplifier is not referred to ground but to a settable voltage ($V_{Tr}$), controlled by a 8-bit potentiometer. As shown in section \ref{sec:VoltageDrift}, this component can offset the positive node to compensate for offsets or thermal drifts of the voltage generation chain. An output voltage divider attenuates the differential signal by a factor of $\varepsilon=R_2/(R_1+R_2)$, down to an energy range that fits the experimental needs (typically from a few keV to several MeV). In a typical application where a heater resistor of few tens of k$\Omega$ is used, an attenuating factor of $\varepsilon=1/201$ is obtained by using $R_1=5$ k$\Omega$ and $R_2=25$ $\Omega$ ultra-stable SMR1DZ-series resistors from Vishay. As a result, the pulse differential amplitude is given by

\begin{equation}\label{eq:VAmplitude}
V_+-V_-=\frac{V_0}{\varepsilon}=\frac{2}{\varepsilon} \left(\frac{R_{D1}}{R_{DAC}}V_{{REF}}+V_{Tr} \right)
\end{equation}

The differential voltage is connected to the heater through a dual channel bistable relay (TQ2-L2 from Panasonic). %In section \ref{sec:OutputBlock} the design choice of the output voltage divider and its performance in terms of thermal stability are shown in detail.     
Each of the two channels of the DAC is connected to two independent output stages, so that 4 channels are available in total.

\subsection{Output stage}\label{sec:OutputBlock}
The output stage is composed of an attenuator circuit and a pair of coil latching bistable relays ($J+$, $J-$ in figure \ref{fig:Scheme}). The relays connect the pulser to the heater only during the pulse delivery, thus preventing electrical noise or disturbance from injecting power towards the bolometers at any other time.  The attenuating circuit is a voltage divider made of ultra-stable and high precision resistors from Vishay, which attenuates by a factor

\begin{equation}\label{eq:Epsilon}
\varepsilon=\frac{V_+-V_-}{V_2-V_1}=\frac{R_1+R_2}{R_2}
\end{equation}

Particular care must be devoted to this block as any drift in the attenuating factor would deteriorate the overall performance.
Moreover, any disturbance picked up after the attenuation would remarkably deteriorate the energy resolution, as it would be superimposed to a much smaller signal. 

The relay on-resistance contributes to $R_1$. If MOSFET relays were used, although the series resistance amounts only to a few ohms, their typical poor thermal stability (of the order of 10 $\%/^\circ$C) could spoil the overall performance in terms of thermal stability.
This guided the choice towards coil latching relays, which ensure a much lower contact resistor of few tens of m$\Omega$ that can be totally neglected. 

According to the manufacturer, $R_1$ and $R_2$ are quoted to ensure a typical thermal drift of $\frac{\partial R}{R \partial T} = +0.05$ ppm/$^\circ$C in the temperature range of interest ($20^\circ$C$-60^\circ$C) and a long term stability of 50 ppm. In addition, the thermal drift of the attenuating factor is even smaller, since the contributions of the two resistors have opposite sign.
The resulting relative drift of $\epsilon$ versus temperature is then given by

\begin{equation}\label{eq:PartDrift}
\begin{split}
\frac{d\varepsilon}{\varepsilon dT} & =\frac{\partial \varepsilon}{\partial R_1}\frac{\partial R_1}{R_1 \partial T}\frac{R_1}{\varepsilon} + \frac{\partial \varepsilon}{\partial R_2}\frac{\partial R_2}{R_2 \partial T}\frac{R_2}{\varepsilon} \\
 & =  \frac{R_1}{R_1+R_2} \left( \frac{\partial R_1}{R_1 \partial T}-\frac{\partial R_2}{R_2 \partial T} \right)  \\
& \ll \frac{\partial R_1}{R_1 \partial T} \sim 5\cdot10^{-2} \,\, \mathrm{ppm}/^\circ\mathrm{C}
\end{split}
\end{equation}
The thermal drift of the attenuating factor is negligible. The resistor $R_2$ is located after the relay, so that the heaters are referred to ground irrespective the relay status. 

The drawback of using such kind of relays is that some disturbances could be induced at the nearby outputs by the magnetic energy released by the coil during the relay transition. Particular care must be therefore addressed to the circuit design and PCB layout in order to minimize the disturbances picked up from the magnetic flux variation. For instance, the same attenuating factor could have been obtained by locating a single ultra-stable resistor $R=2R_2$ across the outputs and using larger value and inexpensive resistor to refer each node to ground when the relays are opened.
At first glance, this solution would appear promising and cheaper. But a loop with a total resistance of $R+2R_2$ would be present at the output, capable to pick up magnetic flux variation. In particular, high frequency electromagnetic fields due to relay transitions, clocks or environmental interference are deleterious as they can induce large flux variation even in a small loop. The noise picked up would then be driven to the heaters, injecting power proportional to the RMS voltage noise amplitude. %This effect would be the most critical factor affecting the stability and the energy resolution of the whole system. 
Conversely, the use of two voltage dividers for each output, placed close to each other, with minimal area and proper relative orientation, composes a pair of identical loops picking up similar noise on both outputs. The power of such signal is consumed to ground through $R_2$ as it acts as common mode signal across the heater. As a result, the symmetric solution proposed in figure \ref{fig:Scheme} turns out to be extremely effective in reducing the power injection due to noise.

\section{Thermal stability}\label{sec:Stab}
The thermal drift is the most important parameter to take into account to ensure the pulser stability. %For example, the CUORE Faraday cage\cite{FaradayCage} hosting the pulser boards is not equipped with any thermalizing system in order to avoid the injection of low frequency disturbances. 
%Slow temperature drifts are expected in the experimental operation and the pulser operation should be insensitive to them. 
According to equations \ref{eq:Joule} and \ref{eq:VAmplitude}, three pulser parameters play an important role in the stability of the energy delivered: the pulse width ($T$), dependent on the on-board reference oscillator, the voltage level produced at the DAC output ($V_0$) and the attenuation factor of the output divider ($\varepsilon$). In particular, the energy drift with respect to a temperature variation is given by differentiating equation \ref{eq:Joule}:
\begin{equation}\label{eq:TotalDrift}
\frac{dE}{EdK}=2\frac{\partial V_0}{V_0\partial K}+\frac{\partial T}{T\partial K}-2\frac{\partial\varepsilon}{\varepsilon \partial K}
\end{equation}
where $K$ is the operating temperature. As described in section \ref{sec:OutputBlock}, the thermal drift of the voltage divider is negligible ($\frac{\partial\varepsilon}{\varepsilon \partial K}\sim 5\cdot10^{-2} \,\mathrm{ppm}/^\circ\mathrm{C}$). The contribution of the pulse width and amplitude are presented in section \ref{sec:ClockDrift} and section \ref{sec:VoltageDrift}, respectively. Each pulser board undergoes a thermal calibration procedure to mitigate the thermal drifts so that $\frac{dE}{EdK}\sim0.1$ ppm$/^\circ$C is achieved in a temperature range going from 20$^\circ$C up to 60$^\circ$C. The method developed to enhance the thermal stability and the obtained results will be shown in section \ref{sec:TotalDrift}.

\begin{figure}[t]
	\centering
		\subfigure[][]{\includegraphics[width=.495\textwidth]{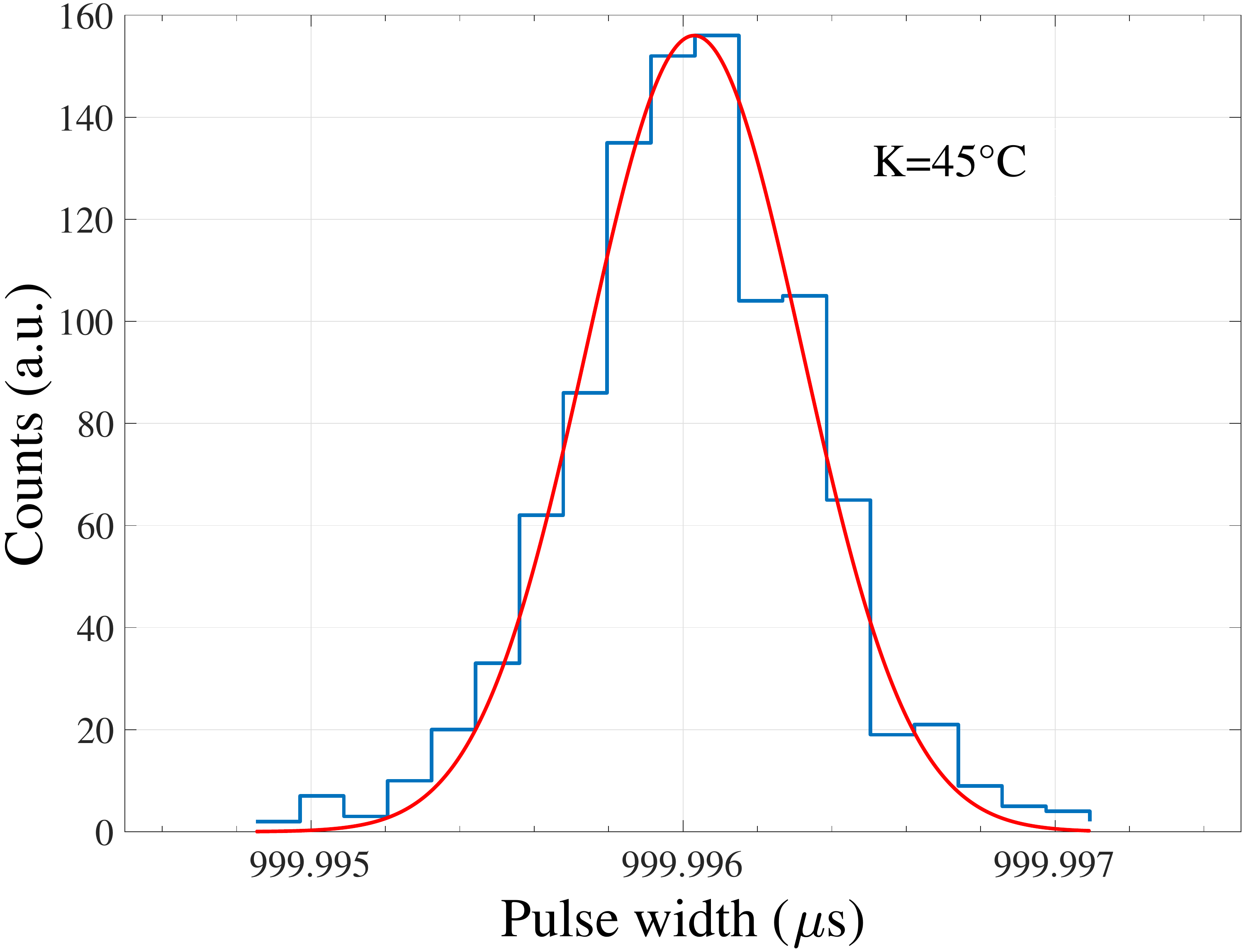}}
		\subfigure[][]{\includegraphics[width=.495\textwidth]{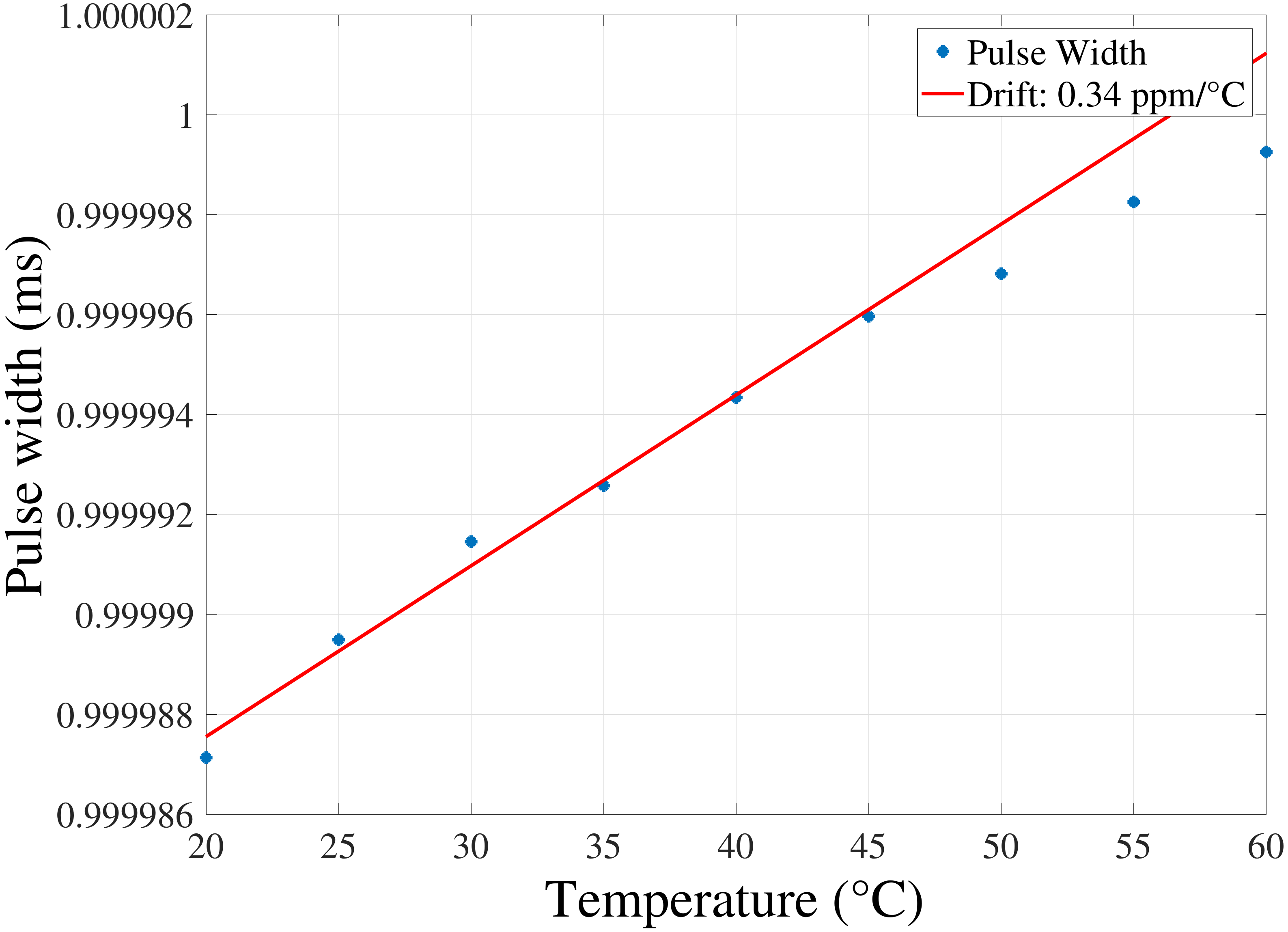}}
		\caption{Pulse width thermal stability: on the left, the pulse width distribution acquired at 45$^\circ$C. The average pulse width is given from the Gaussian center. On the right, the pulse width with respect to temperature. The typical pulse width drift amounts to $0.3-0.4$ ppm$/^\circ$C.}
		\label{Fig:WidthDrift}
\end{figure}

\subsubsection{Pulse width stability}\label{sec:ClockDrift}
The pulse width depends on the performance of the FXO-HC536R reference oscillator. In particular, if the clock period changes as a function of the operating temperature, then the pulse energy drifts proportionally. %The clock reference  drift cannot be directly compensated, but the system is designed to over-compensate the pulse amplitude in order to take it into account. Thus, it is important to characterize the pulse width stability on each board to achieve the best performance. 
The pulse width thermal drift was measured using a Rohde$\&$Schwarz RTO1044 oscilloscope (set with a vertical resolution of 14-bits, 100 MHz bandwidth, time resolution of few tens of picoseconds, dependent on the pulse width), kept at room temperature, while the board under test operated in a climatic chamber from 20$^\circ$C to 60$^\circ$C, with a constant slope of +0.1$^\circ$C per minute. Such slow temperature variation ensures that the pulser board is in thermal equilibrium with respect to the environment during all the measurements. Every 5$^\circ$C, a series of pulses ($V_0= 5$ V, $T= 1$ ms, $\varepsilon=1$) were delivered by the pulser under test. The oscilloscope acquired both positive and negative outputs, calculated the differential signal and measured the pulse width from the delay between the points where the differential signal crosses a threshold placed at half the pulse amplitude on the rising and falling edges. For each temperature of interest, $10^3$ identical pulses were triggered and the measured pulse widths collected into a histogram and fitted with a Gaussian function, as shown in figure \ref{Fig:WidthDrift}.a. The pulse width as a function of temperature could be obtained as the mean value of the Gaussian fit for each sequence of pulses. The typical pulse width thermal drift is shown in figure \ref{Fig:WidthDrift}.b.

The outcomes show that the pulse width slightly increases with temperature. The clock thermal drift is almost linear, with a typical slope of $0.3-0.4$ ppm$/^\circ$C. Only a small deviation from this trend was observed at the highest temperatures ($K\geq50^\circ$C).
The slope of the pulse width with respect to temperature was measured for each device, stored in the on-board EEPROM and used to properly compensate the pulse amplitude.

\subsubsection{Voltage stability}\label{sec:VoltageDrift}

Voltage drifts due to the operating temperature variations can be ascribed to several causes, listed below:
\begin{itemize}
\item \textbf{Voltage reference stability}: the voltage provided by the LTC6655 regulator changes with temperature. This drift is component dependent, non linear and generally shows a minimum drift around $35^\circ$C. The manufacturer quotes the thermal drift to $\pm 2$ ppm$/^\circ$C. Since, for low output signals, only a small fraction of the reference voltage reaches the outputs through the DAC, the thermal drift of the LTC6655 dominates at larger amplitudes.
\item \textbf{DAC stability}: the DAC response depends on the operating temperature in a complex and non linear way. The DAC gain variation with respect to temperature is quoted as $\pm5$ ppm$/^\circ$C, full range scale. This effect dominates the overall voltage instability for small output signals.
\item \textbf{Thermal hysteresis}: the thermal drift of the voltage reference is not only proportional to the actual temperature, but also depends on whether the temperature is increasing or decreasing. This phenomenon is called thermal hysteresis and is particularly deleterious in case of fast temperature changes. In worst case conditions, the thermal hysteresis could result in instability and errors of the order of 30 ppm. Even if the experimental environment is not equipped with an active system to keep the operating temperature stable, the temperature is expected to change slowly during operation, thus mitigating the hysteresis effects.
\item \textbf{Mechanical strains}: temperature increase causes the expansion of all the metals composing the pulser board. Large ground or power supply planes are particularly affected. The mechanical stress caused by the pins soldered to an expanding printed circuit board may cause the output voltage provided by the voltage reference to vary. Proper layout techniques %, such as the omit metallic planes under the voltage reference and tab cut through the PCB around the component, 
significantly reduce the stress on the circuit and mitigate this second order effect.
\item \textbf{Humidity}: with changes in relative humidity, plastic packaging materials absorb water and change the amount of pressure they apply to the die inside. Considering the voltage reference, this strain can cause slight changes in the output of a voltage reference, usually of the order few tens of ppm. To prevent such effect, the voltage reference is packaged in the hermetic LS8 enclosure. Residual second order effects can be ascribed to the mechanical stress applied to the voltage reference by the printed circuit board material which may absorb water.
\item \textbf{Long-term drift}: this non thermal dependent effect regards variations as a function of the operational time. Long-term drift cannot be easily extrapolated from accelerated high temperature testing since this technique tends to give optimistic results. The only reliable way to determine long-term drift is to measure it over the time interval of interest. The pulser board is equipped with voltage references in LS8 package, whose long-term drift is quoted to be $\pm20$ ppm/$\sqrt{\mathrm{kHr}}$. Since this effect is proportional to the square-root of time, its contribution becomes smaller after long periods of operation. %In addition, the periodic calibration with radioactive sources planned during an experiment lifetime are useful to compensate this second order effect.
\end{itemize} 

The voltage drift as a function of temperature has been measured by placing the pulser board in a Vötsch VT-7004 environmental chamber. The temperature is increased linearly from 20$^\circ$C to 60$^\circ$C, with a slope of 0.1$^\circ$C per minute. %Such slow temperature increase ensures that the pulser board is in thermal equilibrium during all the measurement. 
During the temperature cycle, each pulser channel generates a static voltage level (from 1 V to 10 V, 1 V step), measured with a Keithley 3706A Multimeter, kept at room temperature . % Such multimeter is able to measure with extremely high accuracy and stability the voltage delivered from the pulser board, so that even small drifts can be observed.

%\begin{figure}[h!]
%	\centering
%		\subfigure[][]{\includegraphics[width=.495\textwidth]{Immagini/VoltageDrift_2V_Uncompensated.pdf}}
%		\subfigure[][]{\includegraphics[width=.495\textwidth]{Immagini/VoltageDrift_10V_Uncompensated.pdf}}
%		\caption{Intrinsic (non-compensated) pulse amplitude thermal drift. The plot are shown for low (2 V, on the left) and high amplitudes (10 V, on the right).}
%		\label{Fig:VoltageDrift}
%\end{figure}

\begin{figure}[t]
	\centering
		\subfigure[][]{\includegraphics[width=.495\textwidth]{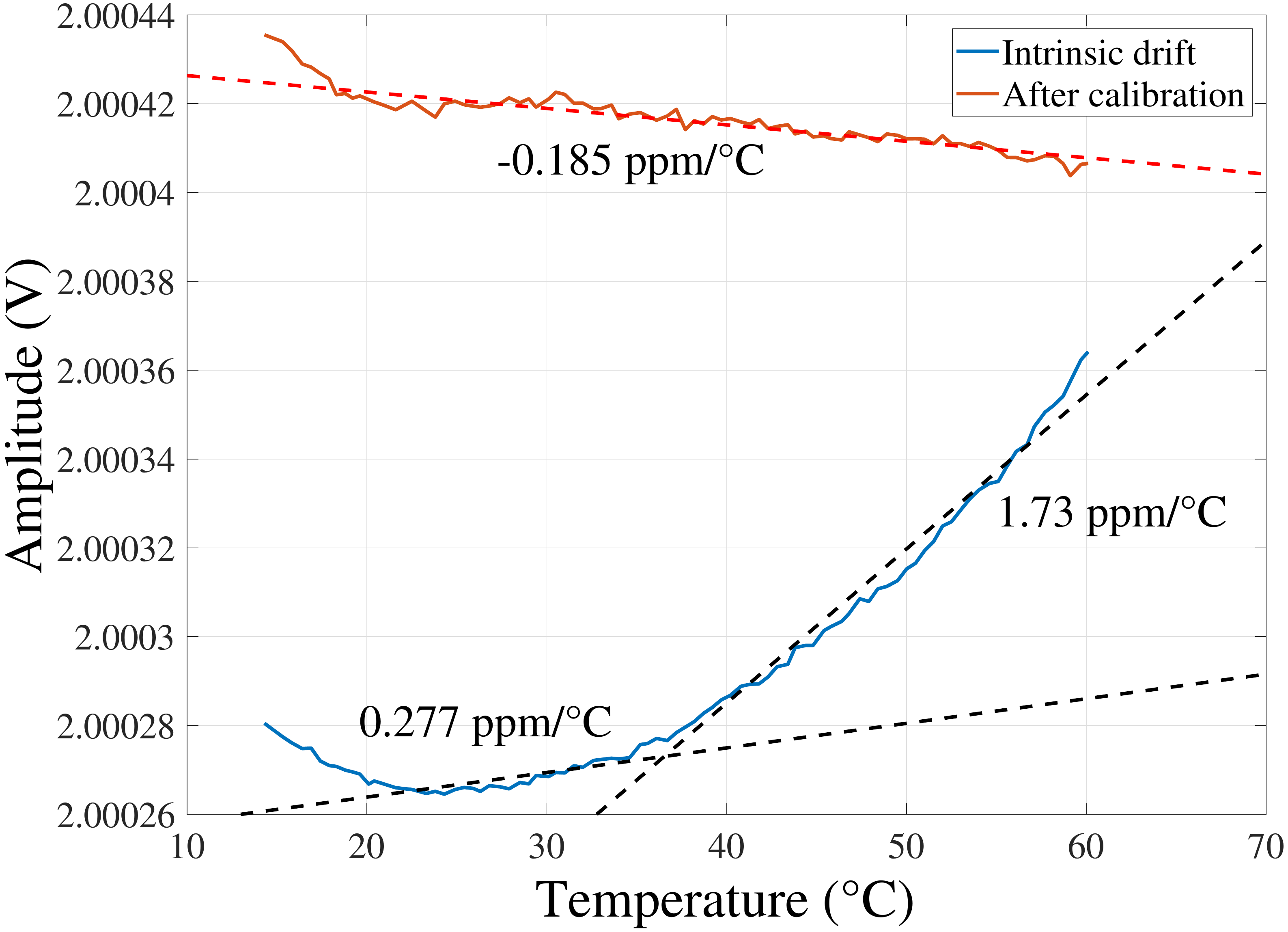}}
		\subfigure[][]{\includegraphics[width=.495\textwidth]{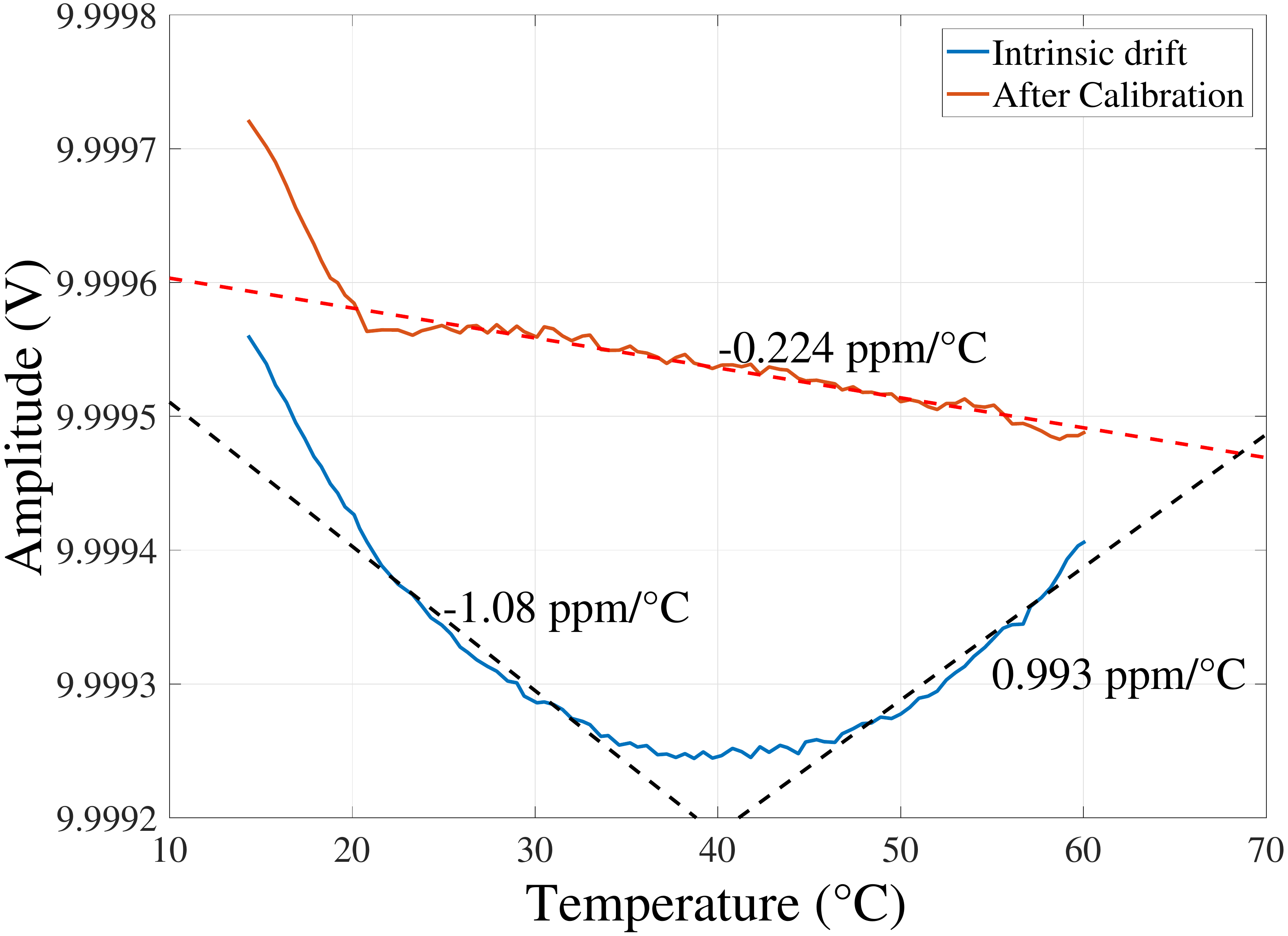}}
		\caption{Intrinsic (non-compensated) and corrected (compensated) pulse amplitude thermal drift, in blue and orange respectively. The plots are shown for low (2 V, on the left) and high amplitudes (10 V, on the right). The drift is purposely over-compensated to correct for the pulse width instability.}
		\label{Fig:VoltageDrift}
\end{figure}

Figure \ref{Fig:VoltageDrift} (blue curves) shows the measured thermal drift as a function of temperature. The drift behaviour significantly changes as a function of amplitude. In particular, at lower amplitudes (figure \ref{Fig:VoltageDrift}.a) the contribution of the voltage reference is largely attenuated and the DAC thermal drift dominates the overall instability. At larger output amplitudes (figure \ref{Fig:VoltageDrift}.b) the relative contribution of the voltage reference increases while the DAC contribution does not change. In this condition the voltage shows a negative slope below $\sim35^\circ$C, while a positive drift at larger temperatures. The typical voltage drift order of magnitude amounts to several ppm$/^\circ$C.

\subsubsection{Total drift}\label{sec:TotalDrift}

Although the thermal drift depends on different effects combined in a complex and non linear way, the intrinsic pulse energy stability is of the order of several parts per million per Celsius degree, as previously described. Nevertheless, most of the residual thermal drift can be compensated by injecting a correcting voltage level ($V_{tr}$) into the positive output buffer (see figure \ref{fig:Scheme} and equation \ref{eq:VAmplitude}). This voltage can be adjusted with a resolution of  $\sim\mu$V by the on-board 8-bit AD5263 digital potentiometer from Analog Devices. Given the complex dependence on temperature of the pulse energy, the drifts must be characterized on each board to be compensated. %Since the thermal drift due to the voltage chain is expected to be very small by design, the attenuation provided by voltage divider is bypassed during the calibration procedure. 
Before undergoing thermal calibration, each board is kept powered for two days at controlled temperature to mitigate the long-term drift contribution. During this period, the digital potentiometer is calibrated and its average step is recorded in the on-board EEPROM. Moreover the pulser baseline is measured and the potentiometer is initially set to compensate the output offset. Baseline offsets lower than 1 $\mu$V are achieved. 
After that, the clock thermal drift is measured for each board in the temperature range from 20$^\circ$C to 60$^\circ$C. The pulse width drift is assumed to be linear and the slope coefficient is measured and stored in the pulser EEPROM memory (see figure \ref{Fig:WidthDrift}.b). Such value is used to properly under-compensate the pulse amplitude in order to correct for the clock thermal drift.
Finally, the voltage thermal drift is measured as explained in section \ref{sec:VoltageDrift} in the temperature range from 20$^\circ$C to 60$^\circ$C, for output levels ranging from 1 V to 10 V, 1 V step. For each amplitude level, the output voltage drift is measured every 5$^\circ$C. Within these intervals the drift is assumed to be linear so that the compensating factor can be extrapolated as a function of the temperature and the desired pulse amplitude.
The boards are also equipped with two diodes which make them able to measure the actual operating temperature from the direct voltage drop of a PN-junction (which typically decreases linearly by $\sim-2$ mV$/^\circ$C). Besides the output drift, also the diodes are calibrated so that the operating temperature can be estimated with a precision of $\sim 1^\circ$C by reading the forward voltage drop by means of the on-board ADC.

\begin{comment}
After the calibration, the pulser board measures the actual temperature before the pulse generation, and sets a proper correcting voltage $V_{tr}$ given by
\begin{equation}\label{VCorr}
\frac{\partial V_{tr}}{\partial K}\Bigr|_{\substack{K_\alpha}} = - \frac{V_0}{2} \left[ 2\frac{\partial V_0}{V_0\partial K}\Bigr|_{\substack{K_\alpha}}+\frac{\partial T}{T\partial K}\Bigr|_{\substack{K_\alpha}} \right]
\end{equation}
where $K_\alpha$ is the actual temperature.
All the factors are known from the calibration procedure.
\end{comment}

After the calibration, the pulser board measures the actual temperature before the pulse generation, and sets a proper correcting voltage $V_{tr}$ chosen so that its added contribution to the voltage drift makes the voltage term in equation \ref{eq:TotalDrift} exactly cancel the pulse width term (while the term in $\epsilon$ is negligible).
Figure \ref{Fig:VoltageDrift} shows superimposed the intrinsic and compensated voltage thermal drift obtained for both low (figure \ref{Fig:VoltageDrift}.a) or high (figure \ref{Fig:VoltageDrift}.b) output levels. Note that the voltage drifts is deliberately undercompensated, so that the effect of the positive linear pulse width thermal drift is also compensated. As expected, the thermal compensation is effective in the temperature range from 20$^\circ$C to 60$^\circ$C. The vertical separation between the two curves is adjusted to compensate any initial offset so to keep the zero-amplitude level as close as possible to the ground reference.

 \begin{figure}[t]
	\centering
		\subfigure[][]{\includegraphics[width=.495\textwidth]{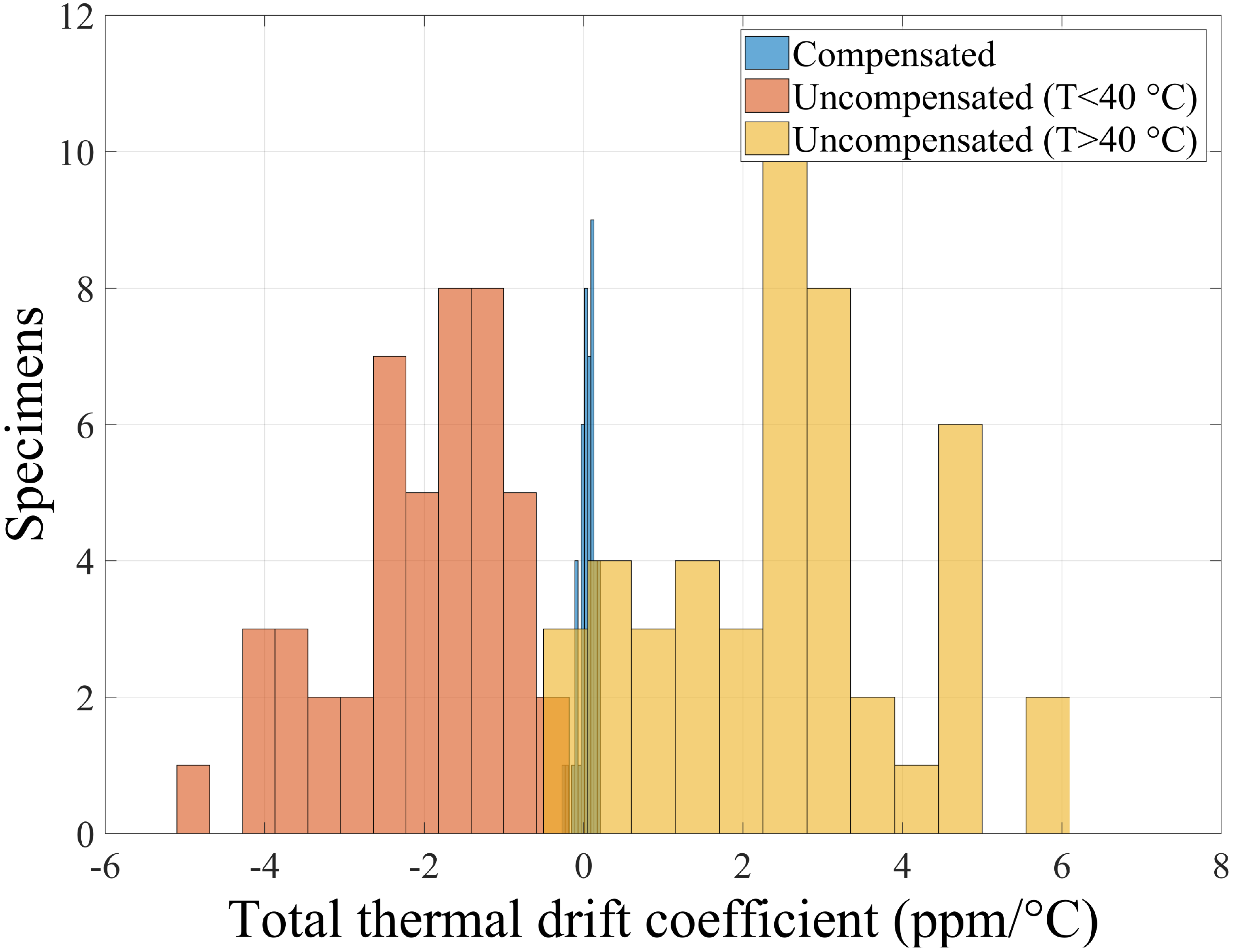}}
		\subfigure[][]{\includegraphics[width=.495\textwidth]{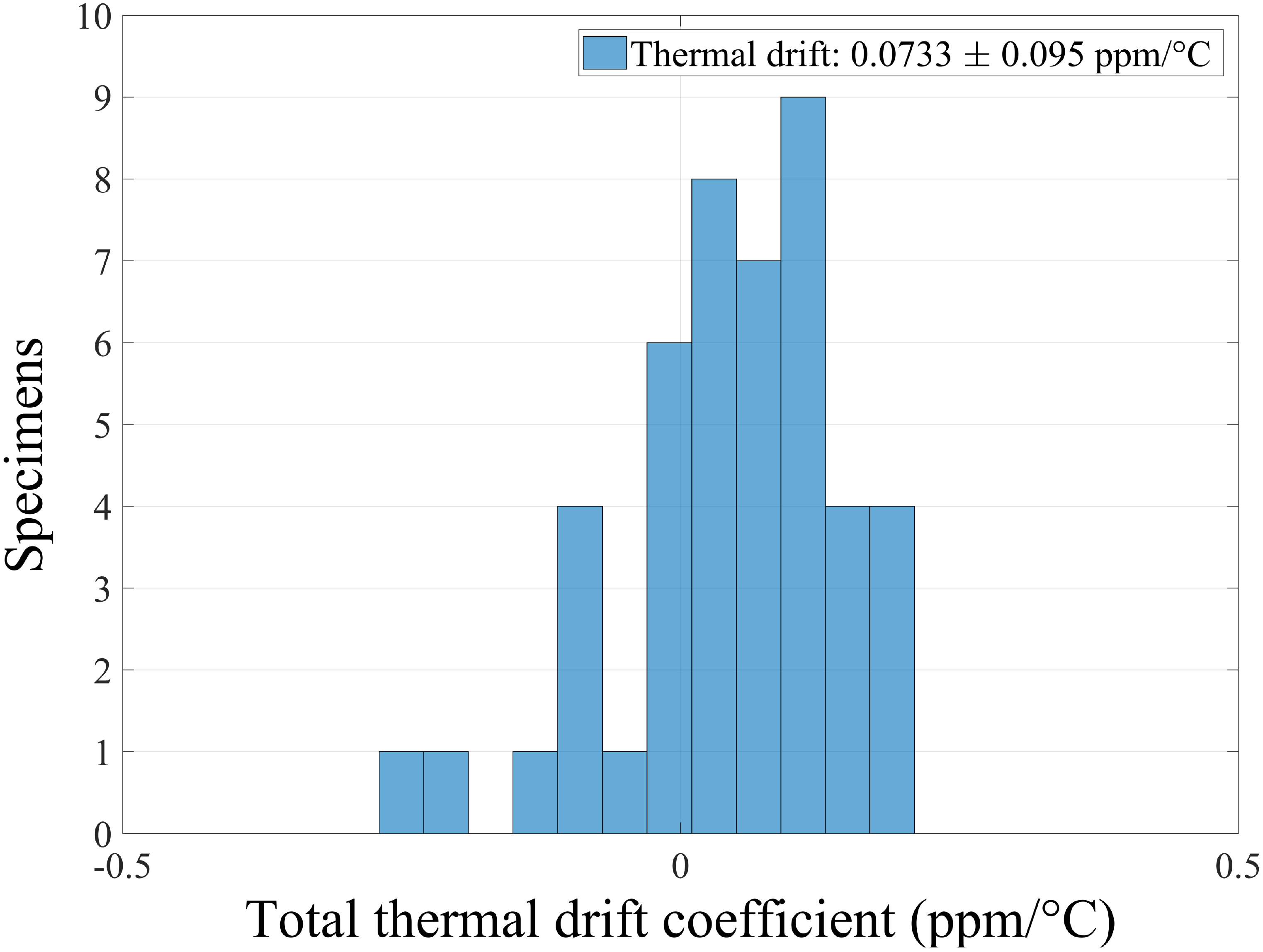}}
		\caption{Efficacy of the thermal drift correction on the whole production lot of pulser boards. Plot (a) shows the superposition of the intrinsic thermal drift distribution at $T<40^\circ$ C (orange histogram) and $T>40^\circ$ C (yellow histogram). The blue histogram is the residual thermal drift distribution after compensation. Plot (b) shows the same result after compensation on a different scale. The residual drift is of the order of 0.1~ppm/$^\circ$C.}
		\label{Fig:Compensation2}
\end{figure}

The calibration procedure allows to reduce the thermal drift of the pulse energy by an order of magnitude, i.e. at the level of $\sim 0.1$ ppm$/^\circ$C. Figure \ref{Fig:Compensation2} shows the residual thermal drifts for a batch of 46 devices. Since the contribution of the voltage reference changes its polarity around 35-40$^\circ$C, the intrinsic thermal drift coefficient distribution is shown considering separately the results obtained at lower or higher temperature range. As can be observed, the intrinsic thermal drift coefficient is $\sim \pm 5-6$ ppm$/^\circ$C. The residual thermal drift coefficient is $\sim \pm 0.1$ ppm$/^\circ$C, almost two orders of magnitude lower. %Finally note that such compensation includes the pulse width and temperature effect but does neglects the voltage divider effect (of the order of $0.3$ ppm$/^\circ$C), long-term drift, thermal drift hysteresis, humidity and mechanical strain.  

\section{Pulse accuracy}\label{sec:Accuracy}

Beside the calibrating pulse stability, the pulser must maximize the reproducibility of the calibrating pulses.  Indeed, considering a sequence of identical pulses, any amplitude or width fluctuation with respect to their nominal value would result in a random variation of the energy delivered to the bolometers and, consequently, in the deterioration of the energy resolution of the calibrating peak in the acquired energy spectrum.
%In principle, this effect can be mitigated by sending a large number of calibrating pulses and taking the average energy provided by each pulse. However, a similar procedure is clearly troublesome and, for practical reasons, cannot be pursued in a large scale experiments. On the other hand,
If the fluctuation of the energy supplied by identical calibrating pulses is much lower than the intrinsic bolometer resolution (a few keV), then the pulser board contributes negligibly to the energy resolution of the peak. %In this case, the energy resolution of the calibrating peak would allow to evaluate the intrinsic resolution of the detector and thus similar to the detector baseline resolution.
%As mentioned, the squared calibrating pulse provided by the pulser board delivers a thermal signal to the bolometers whose energy can be calculated from the Joule formula (\ref{eq:Joule}). 
Concerning the contribution of the pulser board, the fluctuation of the delivered energy is due to the quadratic sum of the contributions from the independent stochastic variations of the pulse amplitude or width.
% as expressed by
%\begin{equation}\label{eq:Resolution2}
%\left(\frac{\sigma_E}{E}\right)^2=\left(\frac{\sigma_E}{E}\right)_{V_0}^2+\left(\frac{\sigma_E}{E}\right)_T^2
%\end{equation}
The contribution of these two sources of uncertainty will be studied separately in the next sections.
As will be described, the contribution of the electrical noise gives the dominant contribution to energy resolution, while the fluctuation of the pulse width is negligible. In particular, it will be shown that $\sigma_E / E \sim 10^{-2} / \sqrt{E[\mathrm{eV}]}$, whatever combination of T and $V_0$ is used.

\subsubsection{Electric noise effects}

\begin{figure}[t] 
\centering
  \def\svgwidth{0.9\columnwidth}
  \graphicspath{{Immagini/}}
  \includesvg{Immagini/SchemaNoise}
  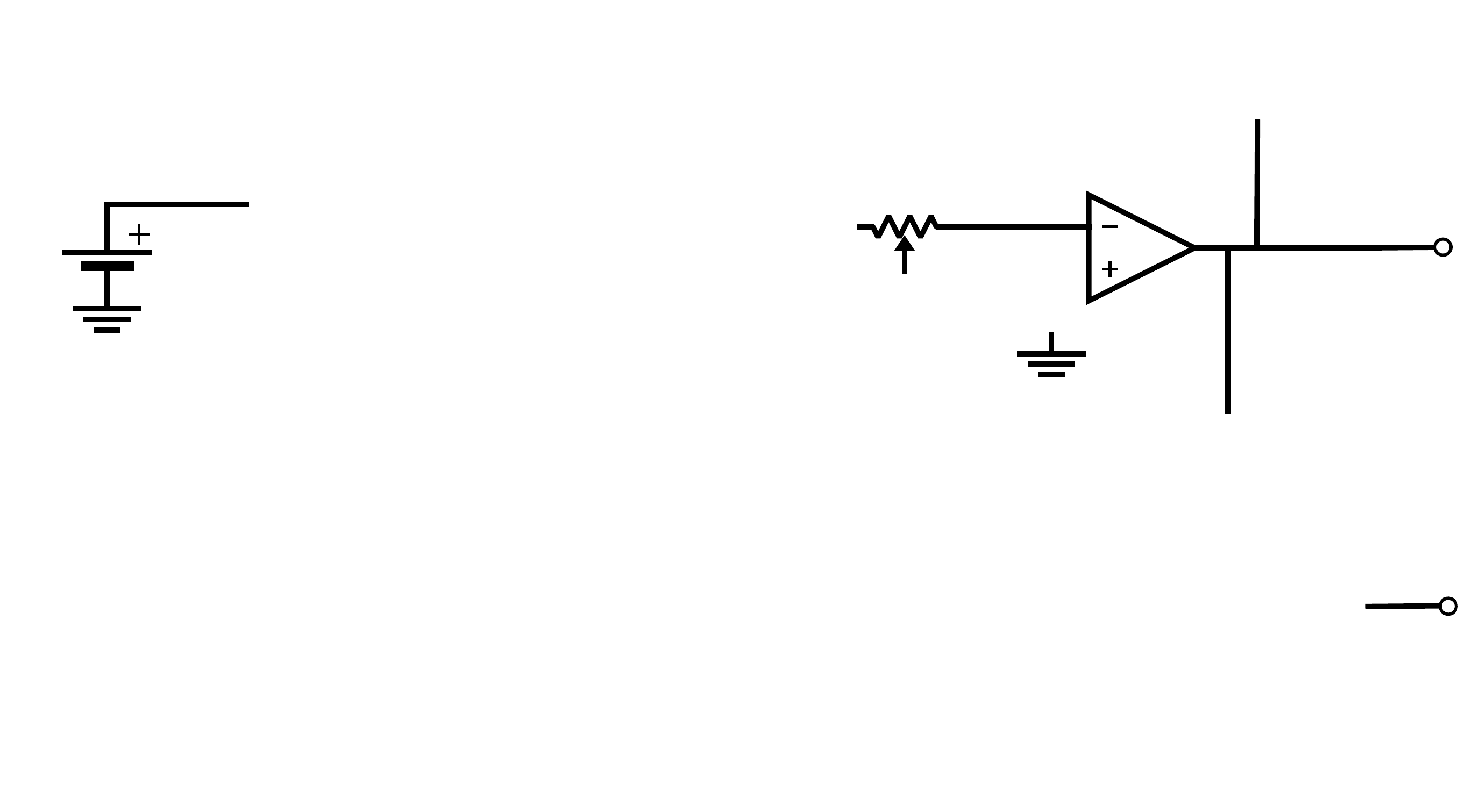%
  \caption{Simplified schematic of a pulser board channel. The most important noise sources are shown.}
  \label{fig:NoiseSource} 
\end{figure}

%The main source of energy fluctuation can be ascribed to the electric noise of the pulser board. %Since the effects of the electric noise on the edges of the pulse will be  studied in the next section, the following dissertation is based on the assumption that the calibrating pulse is ideally rectangular. 
Figure \ref{fig:NoiseSource} shows the most important sources of voltage noise that have to be taken into account.
The first contribution is due to the 5 V reference regulator ($e_{REG}$) and the low noise, high precision, OPA140 operational amplifier ($U_1$) buffering this reference voltage ($e_\pm$). As mentioned, two LTC6655 linear regulator are used in parallel to reduce their noise contribution by $\sim30\%$, so that $e_{REG}\sim 60$ nV/$\sqrt{\mathrm{Hz}}$. The reference voltage is filtered by a RC low-pass filter ($R=50$~$\Omega$, $C=10$~$\mu$F, cut-off frequency $f\sim 300$ Hz) which makes the reference white noise negligible above a few hundreds Hz. Above the cutoff frequency, the noise figure is dominated by the buffer contribution ($e_+=e_-=5$ nV/$\sqrt{\mathrm{Hz}}$). At the negative output, the noise of the reference block ($e_{REF}$) depends on the $R_{DAC}$ adjustable resistor (from 10 k$\Omega$ up to an open circuit) or, equivalently, to the output voltage level $V-$, as expressed by:
\begin{equation}\label{eq:NoiseRef}
e_{REF}\Bigr|_{\substack{V-}} = \left\{ \begin{array}{ll}
\frac{|V_-|}{V_{REG}}\sqrt{e^2_++e^2_-+e_{REG}^2}= 0 \div 60 \,\, \mathrm{nV}/\sqrt{\mathrm{Hz}} & \textrm{if $f \ll 300$ Hz}\\
\frac{|V_-|}{V_{REG}}\sqrt{e^2_++e^2_-}= 0 \div 7 \,\, \mathrm{nV}/\sqrt{\mathrm{Hz}} & \textrm{if $f \gg 300$ Hz}\\
\end{array} \right.
\end{equation}

\begin{figure}[t]
	\centering
		\subfigure[][]{\includegraphics[width=.495\textwidth]{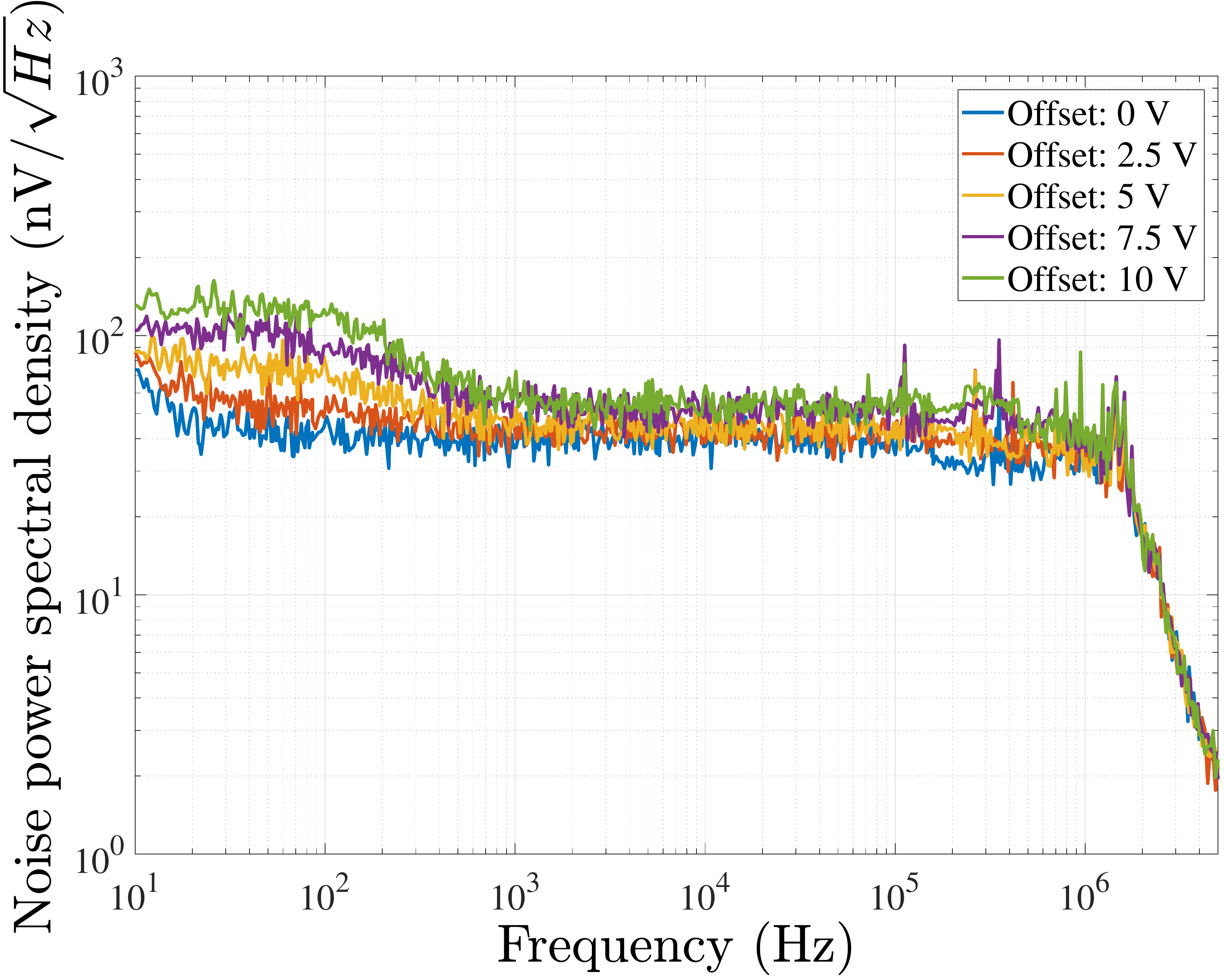}}
		\subfigure[][]{\includegraphics[width=.495\textwidth]{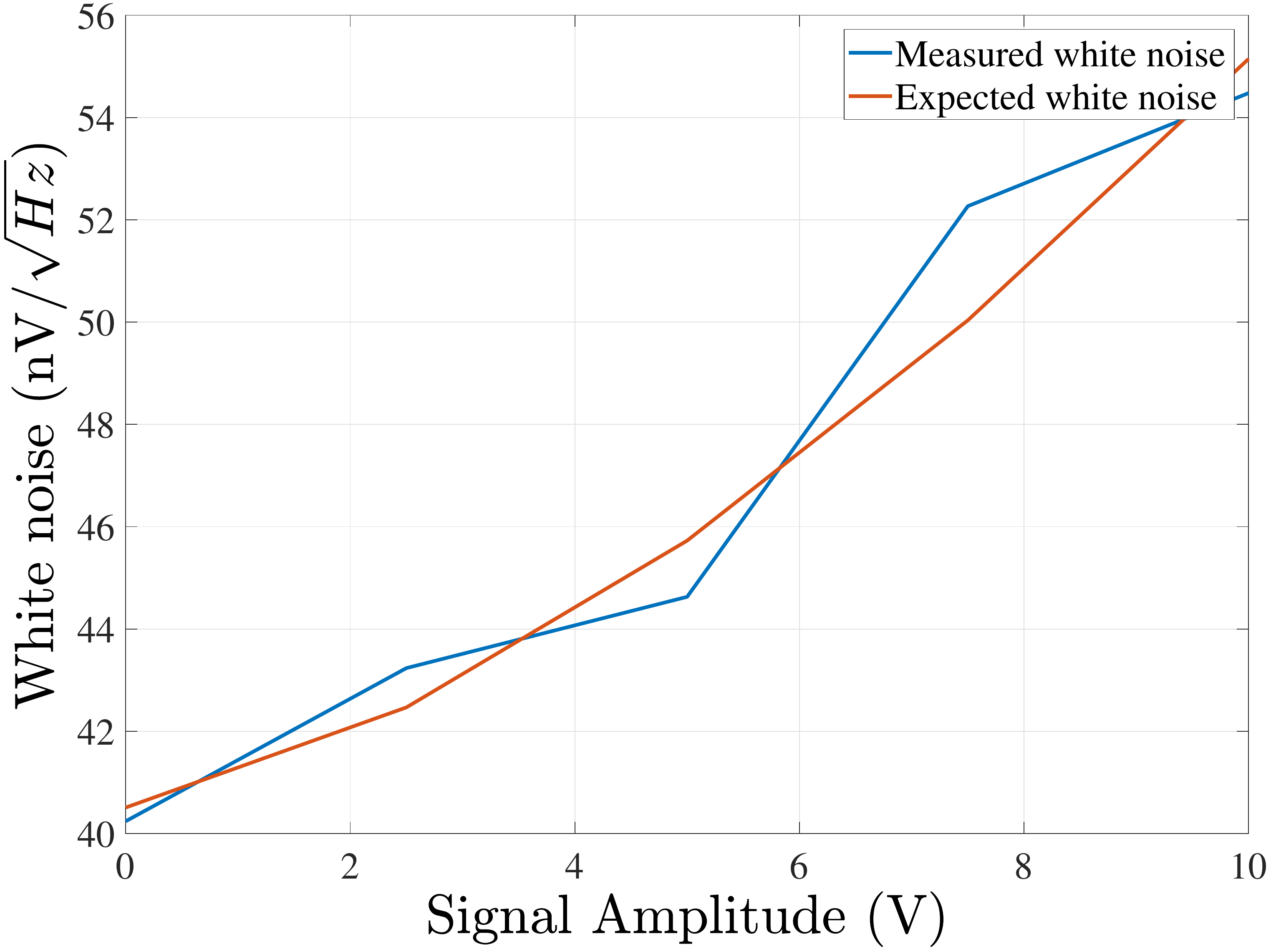}}
		\caption{On the left, noise spectral density at different amplitude settings, measured between the $V_+$ and $V_-$ terminals. On the right, the expected and measured white noise as a function of the signal amplitude.}
		\label{Fig:PosNegBuffer}
\end{figure}

The second noise contribution is ascribed to the negative output buffer stage, due to the Johnson noise of the $R_{DAC}$ and $R_{D1}=10$ k$\Omega$ resistors and the input series noise of the OPA140 operational amplifier $U_2$. At $V-$, the negative output buffer noise ($e_{B_-}$) at room temperature, dependent on the output voltage, is given by:
\begin{equation}\label{eq:NegBuffer}
e_{B_-}\Bigr|_{\substack{V-}}=\frac{|V_-|}{V_{REG}} e_{DAC}+ e_{R1}+ \left( \frac{|V_-|}{V_{REG}} +1 \right) \sqrt{e^2_++e^2_-}= 14 \div 22 \,\, \frac{\mathrm{nV}}{\sqrt{\mathrm{Hz}}}
\end{equation}
Since the noise contributions of the voltage reference and the negative buffer blocks are uncorrelated, the total expected noise at the output is $e_{V_-}=\sqrt{e_{REF}^2+e_{B_-}^2}$, given by equations \ref{eq:NegBuffer} and \ref{eq:NoiseRef}.
The noise at the negative output is injected at the input of the inverting positive output buffer stage. In addition, the positive output buffer stage contributes to the total noise with the Johnson noise of the $R_{D2}=R_{D3}=20$ k$\Omega$ resistors and the input series noise of the OPA140 operational amplifier $U_3$. Omitting the amplitude-dependent contribution coming from the negative output, the positive output buffer noise ($e_{B_+}$) at room temperature referred to the positive output is given by:
\begin{equation}\label{eq:PosBuffer}
e_{B_+}\Bigr|_{\substack{V+}}=\sqrt{e^2_{R2}+e^2_{R3}+8e^2_{+}}= 29 \,\, \frac{\mathrm{nV}}{\sqrt{\mathrm{Hz}}}
\end{equation}
Let us calculate the differential noise expected across the outputs. Since the negative output voltage is driven at the input of the inverting positive buffer stage and affects both output nodes, its weight is doubled. On the other hand, $e_{B_+}$ is an independent source of noise, and has to be quadratically summed to the noise at the negative node. The differential white noise across the heating resistor is described by:
\begin{equation}\label{eq:PosNegBuffer}
e_W=\sqrt{e_{B_+}^2 + 4 \left( e_{REF}^2+e_{B_-}^2 \right)}= \left\{ \begin{array}{ll}
40\div130 \,\, \mathrm{nV}/\sqrt{\mathrm{Hz}} & \textrm{if $f \ll 300$ Hz}\\
40\div55 \,\, \mathrm{nV}/\sqrt{\mathrm{Hz}} & \textrm{if $f \gg 300$ Hz}\\
\end{array} \right.
\end{equation}
The differential noise $e_W$ was measured by amplifying it with a low noise OP27 operational amplifier (close-loop gain of 20 V/V) and acquiring it with an Agilent 4395A spectrum analyser in the frequency range 10 Hz to 5 MHz.
%This amplification stage, coupled in AC through a 2200 $\mu$F capacitor, is used to increase the noise at the output of the pulser board with respect to that due to the spectrum analyzer and the OP27 amplifier.
%The acquired power noise spectrum referred to the pulser output node can be easily calculated by dividing the result by the close-loop gain.
Figure \ref{Fig:PosNegBuffer}.a shows the noise power spectra distribution measured at various output static values. Figure \ref{Fig:PosNegBuffer}.b compares the expected and measured white noise above 300 Hz as a function of the signal amplitude. The expectations are in good agreement with the measurement.

Let us consider how the voltage noise contributes to the thermal power induced towards the bolometers.
The detailed calculations are carried out in appendix \ref{sec:appendix}.
The average injected power is given by
\begin{equation}
\overline{E[eV]}= \frac{\eta}{e R \epsilon^2} \left[
 T V_0^2 +T N_{RMS}^2 (V_0) + \left( \tau-T \right) N_{RMS}^2(0) \right]
 \label{eq:energy1}
\end{equation}
where $T$ is the duration of the square voltage pulse of amplitude $V_0$, and $\tau > T$ is the duration of the entire window, before and after the pulse, where the relays are closed.
Tipically $\tau \simeq$10~ms and $T \simeq$~100~$\mu$s~$\div$~1~ms.
The first term corresponds to the energy injected by an ideal noiseless pulse, as expressed by equation \ref{eq:Joule}.
The quantities $\eta$, $e$, $R$, $\epsilon$ were defined there.
The second and third terms give a small offset related with RMS noise.
There are two RMS noise terms because the output noise depends on the output voltage setting, as was described.
In a practical situation, $V_0 \simeq$~1~V and $N_{RMS}(V_0) \simeq N_{RMS}(0) \simeq$~100~$\mu$V, therefore $N_{RMS}^2/V_0^2 \sim 10^{-8}$.
The noise terms can then be neglected at the level of $10^{-2}$~ppm.

The fluctuation in injected energy is expressed by (again, calculations in appendix \ref{sec:appendix}):
\begin{equation}
\sigma_E^2 [eV^2] \simeq \frac{4}{\pi} \frac{\eta}{e R \epsilon^2} e_W^2 (V_0) \overline{E [eV]}
\label{sigmaE2}
\end{equation}
where $e_W$ is the voltage noise spectral density, assumed white.
Equation \ref{sigmaE2} states that $(\sigma_{E}/\overline{E})^2 \propto \overline{E}^{-1}$, thus the relative resolution is inversely proportional to the square root of the pulse energy and the signal to noise ratio increases at higher energies. Moreover, in this approximation, the pulse resolution does not depend on the pulse amplitude or width, whatever combination of $V_0$ and $T$ has been chosen. By replacing the typical values ($\eta\sim0.5$, $R\sim10$ k$\Omega$, $\varepsilon\sim200$, $e=1.6\cdot 10^{-19}$ C and $e_W\sim50$ nV$\sqrt{\mathrm{Hz}}$) in equation \ref{sigmaE2}, one obtains:
\begin{equation}\label{sigmaE3}
\frac{\sigma_E}{\overline{E}}\sim\frac{10^{-2} eV}{\sqrt{\overline{E}}}
\end{equation}
Equation \ref{sigmaE3} implies that the intrinsic resolution of a typical 1 MeV calibrating pulse amounts to about 20 eV FWHM. This value must be compared to the energy resolution achieved by the detectors.
Typically, in experiments using macro-bolometers such as CUORE or CUPID-0, the target energy resolution is of the order of few keV FWHM, so that the contribution of the pulser board is totally negligible.
The low noise level allows to use the board also for low energy calibration of higher resolution detectors.
The resolution at 10~keV, for instance, is 2 eV FWHM, still lower then the typical resolution of microbolometers for X-ray spectroscopy.

\subsubsection{Time fluctuation effects}

The time fluctuation $\sigma_T$ of the pulse width also contributes in deteriorating the resolution of the generated pulses. As mentioned in section \ref{sec:PulserDesign}, this signal  is provided by the precise 1 MHz oscillator (FXO-HC536R-1 from FOX Electronics) which feeds a fast NC7SV74K8X flip-flop from Fairchild Semiconductor. This architecture ensures both a very precise phase stability, only dependent on the oscillator performance, and fast rise and fall time transitions, limited by the output buffer stage bandwidth, which was set to $\sim1.3$ MHz to ensure high phase margin and low ringing during the pulse transitions.

The time fluctuation has been estimated by acquiring a sequence of 500 identical pulses with the Rohde$\&$Schwarz RTO1044 oscilloscope (4 GHz bandwidth, 10 GHz/Ch sample frequency,  10$^7$ samples per trigger, High Resolution mode with 14 equivalent bits of vertical resolution and bandwidth of 100 MHz). The acquisition window exceeded the pulse width and amplitude by $10\%$, in order to minimize the uncertainties related to the discrete sampling and the limited vertical resolution. The pulses can be triggered when crossing a settable threshold ($V_\alpha$), equal to a desired fraction ($\alpha$) of the pulse amplitude, so that $\alpha=V_\alpha/V_0$. The pulse width is measured as the time difference between when the rising and falling transitions cross half amplitude ($\alpha=0.5$). The observed time fluctuation $\sigma_{T_O}$ is obtained as the standard deviation of the pulse width distribution. 
$\sigma_{T_O}$ is due to two uncorrelated effects:
\begin{itemize}
\item $\sigma_{T_T}$ is directly referred to the uncertainties affecting the time dimension. The first contribution to be taken into account is the phase stability of the reference oscillator ($\sigma_{T_p}$). According to the datasheet provided by the manufacturer, the HC536R-1 ensures a phase stability of $\pm25$ ppm with respect to the clock nominal frequency (1 MHz). The second contribution is the oscilloscope resolution on the horizontal axes (time scale). In particular, the finite record  buffer length and the sample frequency limit the maximum sample frequency to the ratio between the acquiring window width and the memory buffer length. For instance, for pulse widths of $T=500$ ms, the oscilloscope time resolution is $\sigma_{T_{osc}}=$55 ps. Note that such uncertainties affect both transitions of the calibrating pulse, so that the total pulse width fluctuation is given by
\begin{equation}\label{eq:PhaseStability}
\sigma_{T_T} = \sqrt{2 (\sigma_{T_p}^2 + \sigma_{T_{osc}}^2)} \sim 85 \, \mathrm{ps}.
\end{equation}

\item $\sigma_{T_N}$ is related to the electric noise and the finite bandwidth of the output buffer stage. The pulse width is evaluated from the time where the pulse crosses a fixed voltage threshold $V_\alpha=V_0/2$. If the transitions were ideally vertical (infinite pulser bandwidth), then the voltage noise would not affect the measurements. On the contrary, given a finite bandwidth, the trigger time fluctuates as a consequence of the vertical position uncertainty of the trigger point due to the electric noise. This effect stochastically affects both rising ($\sigma_{T_R}$) and falling ($\sigma_{T_F}$) pulse edges, as expressed by
\begin{equation}\label{eq:JitterNoise1}
\sigma_{T_N}^2=\sigma_{T_R}^2\Bigr|_{\substack{\alpha}}+\sigma_{T_F}^2\Bigr|_{\substack{\alpha}}=\left( \frac{dV_R}{dt} \right)^{-2} \Bigr|_{\substack{t_\alpha}}\sigma_{V}^2+\left( \frac{dV_F}{dt} \right)^{-2} \Bigr|_{\substack{t_\alpha}}\sigma_{V}^2
\end{equation}
where $t_\alpha$ is the trigger time at the trigger threshold and $\sigma_{V}$ is the RMS noise at the oscilloscope input. For a finite bandwidth $BW=1/(2\pi\tau)=1.3$ MHz, the rising and falling edges can be expressed as:
\begin{equation}\label{eq:DerTimeNoise}
\begin{split}
V_R(t)&=V_0\,(1-e^{-t/\tau})\\
V_F(t)&=V_0\,e^{-t/\tau}
\end{split}
\end{equation}
with $\tau\sim250$ ns. By using equation \ref{eq:DerTimeNoise}, one can solve equation \ref{eq:JitterNoise1} and obtain:
\begin{equation}\label{eq:JitterNoiseAlpha}
\sigma_{T_N}^2\Bigr|_{\substack{\alpha}}= \left[ \frac{(1-\alpha)^2+\alpha^2}{(1-\alpha)^2\alpha^2} \right] \frac{\tau^2}{V_0^2} \, \sigma_{V}^2
\end{equation}

$\sigma_{T_N}^2|_\alpha$ has a minimum for $\alpha=0.5$. In this case, equation \ref{eq:JitterNoiseAlpha} becomes:

\begin{equation}\label{eq:JitterNoiseAlpha1}
\sigma_{T_N}^2\Bigr|_{\substack{\alpha=0.5}}= 8 \, \frac{\tau^2}{V_0^2} \, \sigma_{V}^2
\end{equation}

As mentioned, $\sigma_{V}^2$ is the RMS electrical noise read by the RTO1044 oscilloscope. Thus:

\begin{equation}\label{eq:JitterNoiseAlpha1111}
\sigma_{V}^2=\sigma_{V_{RMS}}^2+\sigma_{V_{pre}}^2+\sigma_{V_{dig}}^2
\end{equation}

where $\sigma_{V_{RMS}}\sim 100$ $\mu$V is the pulser RMS noise calculated over the full-bandwidth of the output buffer, $\sigma_{V_{pre}}$ is the RMS noise of the oscilloscope input amplifier and $\sigma_{V_{dig}}$ is the RMS voltage resolution of the oscilloscope ADC. The two latter contributions are calculated over the full oscilloscope bandwidth (up to 100 MHz) . Note that $\sigma_{V_{RMS}}$ slightly increases with $V_0$  (as shown in the previous section),  $\sigma_{V_{dig}}=\gamma V_0$ ($\gamma \sim 400$ $\mu$V/V, according to the manufacturer specs) since the equivalent voltage step of the least significant bit is proportional to the vertical scale set up to the oscilloscope, while $\sigma_{V_{pre}}$ is independent of $V_0$ so that its contribution becomes negligible with respect to $\sigma_{V_{dig}}$ at large signal amplitudes. 
\end{itemize}
$\sigma_{T_O}$ is given by the above contributions summed in quadrature. From equation \ref{eq:PhaseStability}, \ref{eq:JitterNoiseAlpha1} and \ref{eq:JitterNoiseAlpha1111} one obtains:
\begin{equation}\label{eq:JitterNoise4}
\begin{split}
\sigma_{T_O}^2&=\sigma_{T_T}^2+\sigma_{T_N}^2\Bigr|_{\substack{\alpha=0.5}}\\
&=2 \sigma_{T_p}^2 + 2 \sigma_{T_{osc}}^2 + \frac{8\tau^2}{V_0^2} \left( \sigma_{V_{RMS}}^2+\sigma_{V_{pre}}^2 \right) + 8\tau^2\gamma^2
\end{split}
\end{equation}
Equation \ref{eq:JitterNoise4} describes the jitter measured at half of the pulse amplitude, univocally defined by $\alpha=0.5$. The first and third terms are due to the pulser board, while the other factors represent the setup sensitivity. %Note that this value is not directly related to an effective pulse width fluctuation, as it concerns the voltage fluctuation of a single point of the pulse edge. 
%The effective impact of such fluctuation to the pulse energy is lower (by a factor $\sim 68\%$ \cite{AProgPS}), since the effects of one point of the pulse edge is partially compensated by the others. 
Note that, $\sigma_{T_O}$ decreases by increasing the signal amplitude and, neglecting all the contributions from the pulser board, has a minimum for $\sigma_{T_O}^2=\sigma_{T_T}^2 + 8\tau^2\gamma^2\sim 280$ ps, dominated by the voltage resolution of the RTO1044 oscilloscope. In such condition, $\sigma_{T_O}$ has to be considered as an upper limit of the intrinsic calibrating pulse fluctuation.

\begin{figure}[t]
	\centering
		\begin{minipage}[t]{.4\textwidth}
			\includegraphics[height=0.75\textwidth]{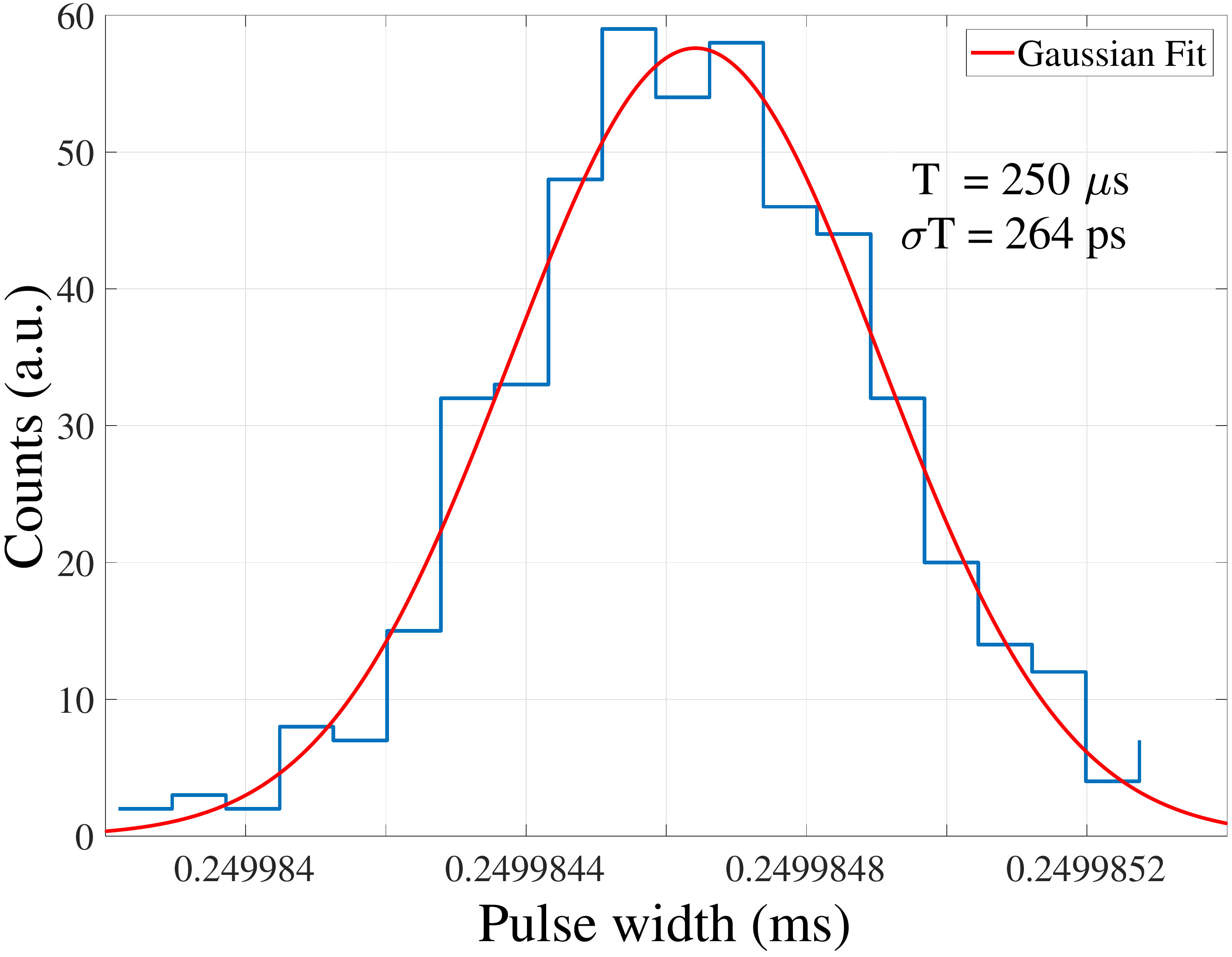}
			\caption{Pulse width distribution acquired recording 500 pulses, 10 V high, 250 $\mu$s wide. In red, the Gaussian function fitting the distribution. The measured average pulse width and its fluctuation are given from the center and standard deviation of the fitting curve.}
			\label{fig:GaussHist}
		\end{minipage}%
	\hspace{50mm}%
		\begin{minipage}[t]{.4\textwidth}
			\includegraphics[height=0.75\textwidth]{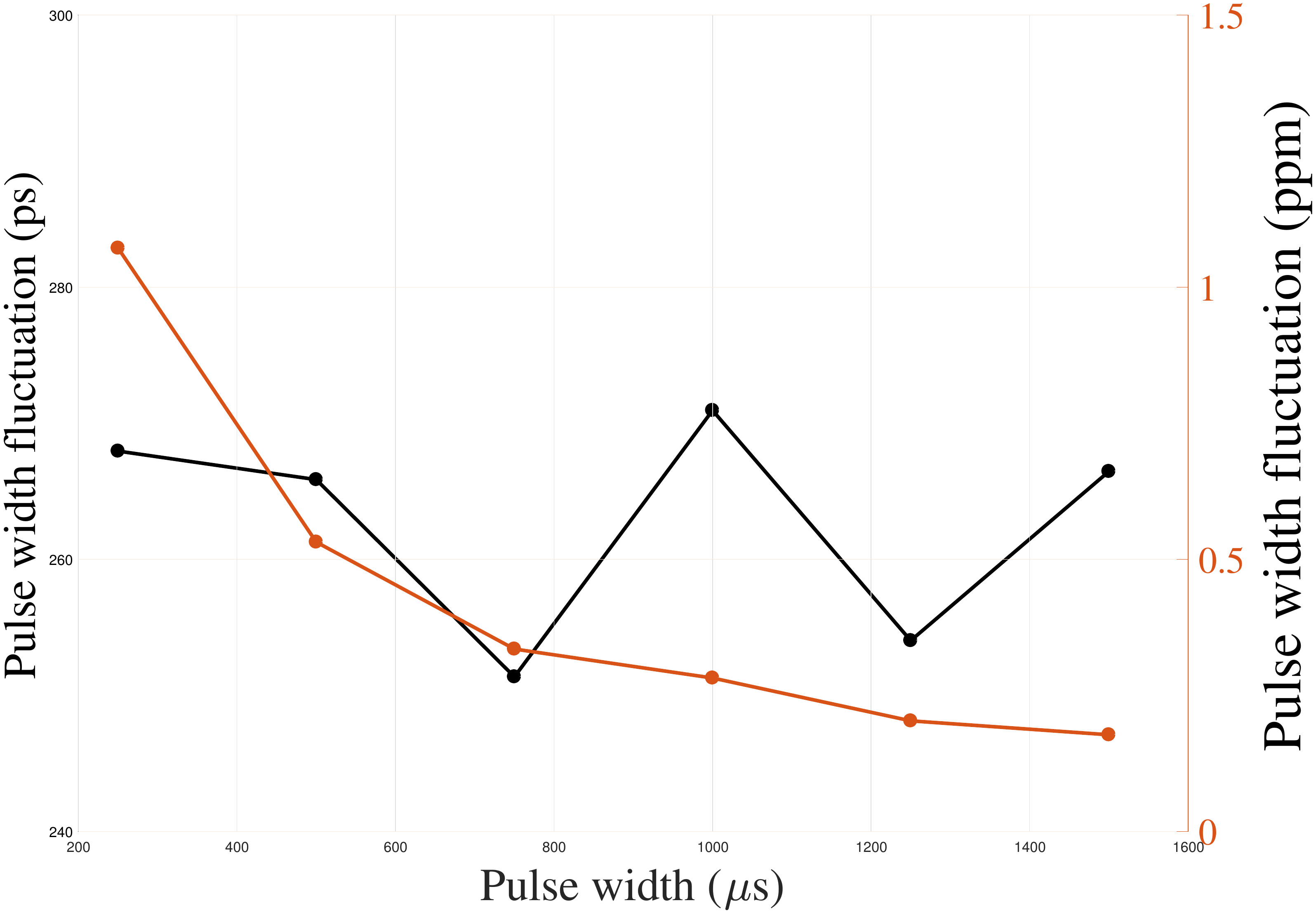}
			\caption{Pulse width distribution (500 pulses, 10 V high) as a function of the pulse width shown in absolute (black curve and left vertical axis) and relative (orange curve and right vertical axis) scale. For signals wider than $\sim 500$ $\mu$s, the fluctuations are below 0.5 ppm.}
			\label{fig:JitterW}
		\end{minipage}%
	\hspace{10mm}%
		\begin{minipage}[t]{.4\textwidth}
			\centering
			\includegraphics[height=0.75\textwidth]{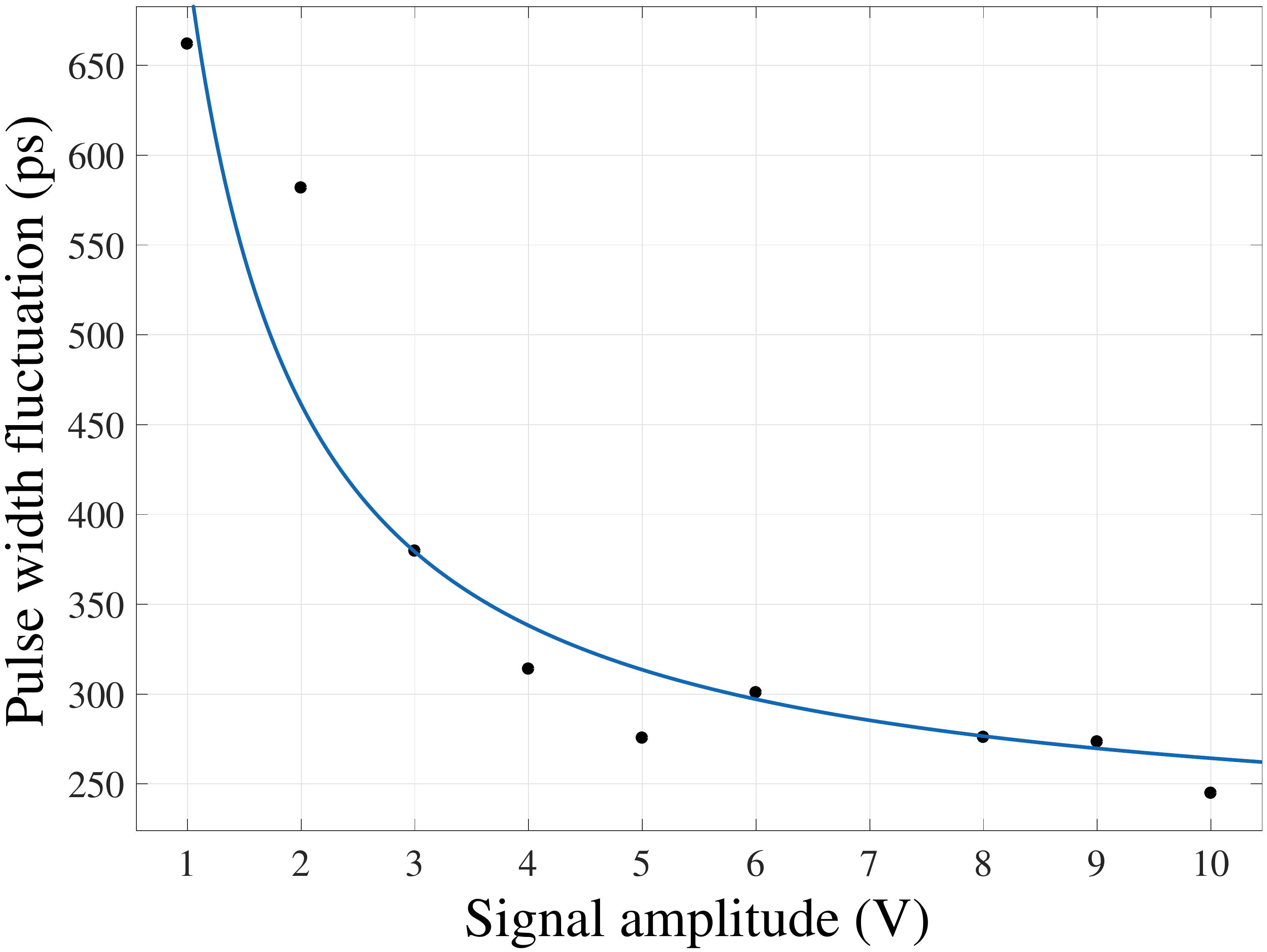}
			\caption{Pulse width distribution (500 pulses, 250 $\mu$s wide) as a function of the pulse amplitude. At larger amplitudes, the relative magnitude of the oscilloscope noise decreases allowing to reach better sensitivities.}
			\label{fig:JitterV}
		\end{minipage}%
\end{figure}

Figure \ref{fig:GaussHist} shows the typical pulse width distribution acquired with the RTO1044 oscilloscope generating a sequence of 500 pulses, 10 V high, 250 $\mu$s wide. The measure was performed keeping the pulser board in a Vötsch VT-7004 environmental camber, at a stable temperature of 20$^\circ$C.  The distribution is well described by a Gaussian function centered at the average measured pulse width, whose standard deviation $\sigma_{T_O}\sim270$ ps describes the observed width fluctuation. Since the normal distribution fits the data, then the pulse width fluctuation is only due to random effects. In particular, neither clock transition miscounting nor missed triggers was observed. This proves the goodness of the design hardware choice and the satisfactory operation of the firmware.  This procedure has been repeated at various pulse widths and amplitudes. The pulse width fluctuation as a function of T and $V_0$ is shown in figure \ref{fig:JitterW} and figure \ref{fig:JitterV}. $\sigma_{T_O}$ remains constant with respect to the pulse width and is close to the setup sensitivity, limited to $\sim 280$ ps by the resolution of the digital conversion. This proves that the pulser contribution is much lower than the measurement sensitivity. The observed pulse width fluctuation as a function of the signal amplitude, shown in figure \ref{fig:JitterV}, decreases inversely proportional to the pulse amplitude until it reaches the best available sensitivity for signals larger than 4 V. The larger time fluctuation at lower amplitude is due to noise effects that are independent of the signal amplitude, such as the electric noise of the oscilloscope front-end or that of the pulser output buffer stage. Nevertheless, the contribution of the oscilloscope at low amplitudes is estimated to be several times larger then that of the pulser board. %A reasonable estimation of the observed  pulse width fluctuation is $\sigma_{T_O} \sim 300$ ps, which represents the upper limit of the effective energy dispersion due to the width fluctuation. 
Considering a typical calibrating pulse of $T=500$ $\mu$s, the expected contribution to the intrinsic calibrating energy resolution is:
\begin{equation}\label{eq:JitterResult}
\left( \frac{dE}{E} \right)_T\leq \frac{\sigma_{T_O}}{T} \sim 1 \, \mathrm{ppm}
\end{equation}
which is an order of magnitude below the fluctuation due to voltage noise, and several orders of magnitude below the typical resolution of bolometric detectors.

\section{Baseline stabilization}\label{sec:PID}

The design of the pulse generation board makes the device able to deliver at its outputs not only a calibrating pulse but also a constant voltage level and, if necessary, to superimpose pulses over it. The capability of providing an adjustable and precise static level can be used to stabilize the temperature of the detectors with respect to low frequency drifts due to the cryogenic setup.
In particular, a settable power can be injected by means of heaters located on the mixing chamber of the cryostat. The consequent temperature increase can be measured by cryogenic thermometers read out by the on-board 24-bit, dual differential channel, AD7732 Analog-to-Digital converter form Analog Devices. This signal is compared to the target temperature and the power is adjusted to minimize the mismatch. % compensate for the low frequency drift. 
Such closed-loop architecture must be customized to take into account the open-loop response of the cryostat. In particular, a proportional-integral-derivative (PID) controller can be implemented in software to manage the feedback mechanism. The pulse generation board, being able to provide power and read out the temperature, represents the core of the PID system.
The PID calculations can be implemented directly in the device firmware, or can be performed by a remote master. 

\begin{figure}[t]
	\centering
	\includegraphics[width=0.8\textwidth]{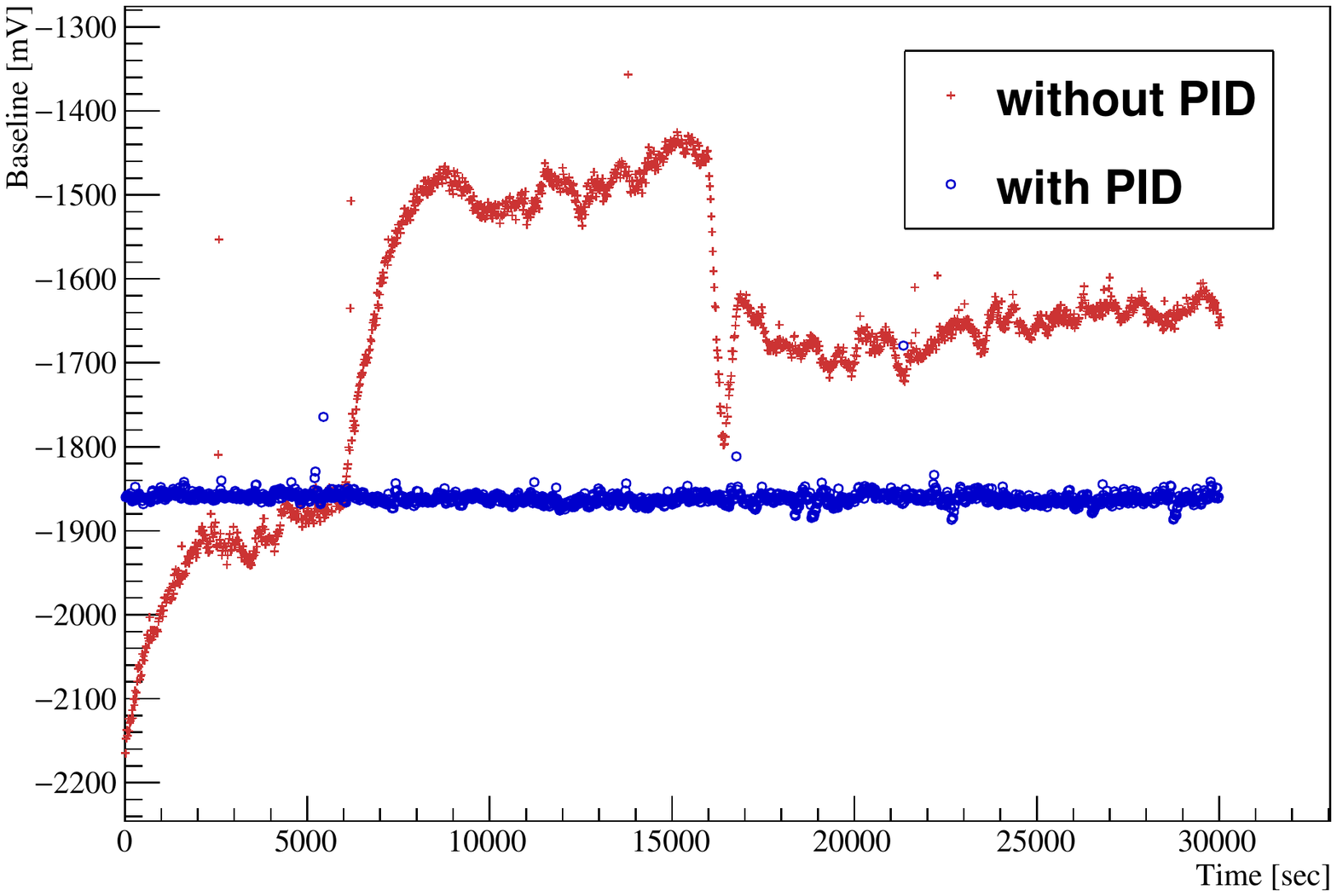}
	\caption{Bolometer baseline acquired in the CUORE experiment with the PID controller activated (blu line) or deactivated (orange line).}
	\label{fig:PID}
\end{figure}

This baseline stabilization technique is currently used in the CUORE and LUCIFER/CUPID-0 experiments. Figure \ref{fig:PID} shows the effectiveness of this approach to compensate the low frequency baseline drift in a CUORE crystal. The orange curve shows the unavoidable instability which usually occurs in the detector baseline in a 9 hour run. Since the baseline drifts are proportional to the crystal temperature variation, they must be continuously monitored and correlated to the detector response to ensure stability and reproducibility over the experiment live time. Superimposed to the orange curve, the blue line represents the baseline of the same detector in a 9 hour run in which the PID controller based on the pulse generation board is used. As can be easily seen, the detector baseline stability is greatly improved and the intrinsic instabilities of the cryogenic system are completely compensated.
Besides baseline stability, the PID controller can be also used to tune the operating temperature of the crystals, if a slightly higher temperature than that offered by the cryostat is required.

\section{Conclusions}\label{sec:Conclusion}
In this paper, a circuit able to provide ultra-stable and ultra-precise pulses has been presented. The device has been specifically designed provide reference thermal signals via Joule effect, to calibrate bolometric detectors. The pulse generation board is currently used in several experiments (CUORE, LUCIFER/CUPID-0, COSINUS, all at LNGS) where it proved to be capable to stabilize the detector response against the instabilities usually affecting such cryogenic measurements. The pulse delivery and all the other features can be fully controlled and monitored remotely through a optically decoupled CAN bus. An additional daisy-chained optical trigger is also available. The design choices have been presented in detail. After undergoing a calibration procedures, a pulse energy stability of the order of 0.1 ppm/$^\circ$C has been achieved in the temperature range of interest (20$^\circ$C - 60$^\circ$C). Moreover, an energy resolution of $\sim$20 eV FWHM has been accomplished for typical 1 MeV pulses, and $\sim$2 eV FWHM for 10 keV pulses, thanks to the low electrical noise and the high timing precision.
The pulser board can also be used in a close-loop architecture (such as PID controllers) to compensate for the slow instabilities of the detector baseline and to optimize the operating temperature to achieve the best signal-to-noise ratio.

\acknowledgments
The CUORE and LUCIFER/CUPID-0 Collaborations thank the Director and staff of the Laboratori Nazionali del Gran Sasso and the technical staff of our laboratories.
This work was supported by the Istituto Nazionale di Fisica Nucleare (INFN); the National Science Foundation under Grant Nos. NSF-PHY-0605119, NSF-PHY-0500337, NSF-PHY-0855314, NSF-PHY-0902171, NSF-PHY-0969852, NSF-PHY-1307204, NSF-PHY-1314881, NSF-PHY-1401832, and NSF-PHY-1404205; the Alfred P. Sloan Foundation; the University of Wisconsin Foundation; and Yale University. This material is also based upon work supported by the US Department of Energy (DOE) Office of Science under Contract Nos. DE-AC02-05CH11231, DE-AC52-07NA27344, and DE-SC0012654; and by the DOE Office of Science, Office of Nuclear Physics under Contract Nos. DE-FG02-08ER41551 and DE-FG03-00ER41138.
This work also received support in the frame of the LUCIFER experiment, funded by the European Research Council under the European Union’s Seventh Framework Programme (FP7/2007-2013)/ERC Grant Agreement no. 247115.
%Authors thanks the Publication Board of the CUORE experiment for the kind review of this paper.

\appendix
\section{Detailed calculations of injected energy and its fluctuation}
\label{sec:appendix}

In this appendix we will calculate the expressions for injected energy and its fluctuation due to electronic noise of the pulser output voltage.
If we generate K identical pulses $V(t)$ with different noise $N_k(V,t)$, the average injected energy is expressed by
\begin{equation}
\overline{E}= \left<E_k \right> =
\frac{\eta}{e R \epsilon^2} \left< \int_{-\tau/2}^{\tau/2} \left[ V(t) + N_k(V(t), t)  \right]^2 dt  \right>
\end{equation}
where the integration extends over the entire period $\tau$ when the relays are closed, and the quantities $\eta$, $e$, $R$, $\epsilon$ were defined in the paper for equation \ref{eq:Joule}.
The angle brackets indicate averaging over $k$:
\begin{equation}
\left<E_k \right> =  \lim_{K\to\infty} \frac{1}{K} \sum_{k=1}^{K} E_k
\end{equation}
By using the fact that $V(t)$ is a square pulse with amplitude $V_0$ and duration $T < \tau$, we obtain
\begin{equation}
\overline{E}= \frac{\eta}{e R \epsilon^2} \left<
 \int_{-T/2}^{T/2} V_0^2 dt + \int_{-T/2}^{T/2} N_k^2(V_0,t) dt
 + 2 \int_{T/2}^{\tau/2} N_k^2(0,t) dt + 2 \int_{-T/2}^{T/2} V_0 N_k(V_0,t) dt \right>
\end{equation}
By averaging over $k$ the fourth term is cancelled, and we obtain:
\begin{equation}
\overline{E}= \frac{\eta}{e R \epsilon^2} \left[
 T V_0^2 +T N_{RMS}^2 (V_0) + \left( \tau-T \right) N_{RMS}^2(0) \right]
 \label{eq:appenergy1}
\end{equation}
That is the expression for average injected energy used in the paper, equation \ref{eq:energy1}.

Let us now consider the fluctuation of the injected energy, expressed by
\begin{equation}
{\sigma_E^2}= \left<\sigma_k^2\right> =
 \left<\left(E_k - \overline{E}\right)^2 \right>
\simeq
4 V_0^2 \left(\frac{\eta}{e R \epsilon^2}\right)^2
\left<
\left[ \int_{-T/2}^{T/2} N_k(V_0, t) dt \right]^2
\right>
\end{equation}
where all the terms with higher powers in $N_k$ were dropped.
This expression can be managed as follows:
\begin{equation}
{\sigma_E^2} =4 V_0^2 \left(\frac{\eta}{e R \epsilon^2}\right)^2
\left<
\left[\int_{-T/2}^{T/2} \left( \frac{1}{2 \pi} \int_{-\infty}^{\infty} N_k(V_0, \omega) e^{i \omega t} d \omega \right) dt \right]^2
\right>
\end{equation}
\begin{equation}
=4 V_0^2 \left(\frac{\eta}{e R \epsilon^2}\right)^2
\left<
\left[ \frac{1}{2 \pi} \int_{-\infty}^{\infty} N_k(V_0, \omega) \left( \int_{-T/2}^{T/2} e^{i \omega t} dt \right) d \omega \right]^2
\right>
\end{equation}
\begin{equation}
=4 V_0^2 \left(\frac{\eta}{e R \epsilon^2}\right)^2
\left<
\left[ \frac{1}{2 \pi} \int_{-\infty}^{\infty} N_k(V_0, \omega) \left[ \frac{2}{\omega} \sin \left(\frac{\omega T}{2} \right) \right] d \omega \right]^2
\right>
\label{eq:appsigma1}
\end{equation}
We can now consider the following approximation:
\begin{equation}
\frac{\sin x}{x} \simeq  \theta \left(\frac{2x}{\pi}+1 \right)- \theta \left(\frac{2x}{\pi}-1 \right)
\end{equation}
where $\theta$ is the Heaviside step function, and the parameters were chosen to have the same value in $x=0$ and the same normalization by integrating over $x$.
By using the approximation, equation \ref{eq:appsigma1} becomes
\begin{equation}
{\sigma_E^2} \simeq 4 V_0^2 \left(\frac{\eta}{e R \epsilon^2}\right)^2
\left<
\left[ \frac{T}{2 \pi} \int_{-\infty}^{\infty} N_k(V_0, \omega)
\left[ \theta \left(\omega + \frac{\pi}{T} \right)- \theta \left( \omega - \frac{\pi}{T} \right) \right] d \omega \right]^2
\right>
\end{equation}
\begin{equation}
= 4 V_0^2 \left(\frac{\eta}{e R \epsilon^2}\right)^2
\left<
 \left[ \frac{T}{2 \pi} \int_{-\pi/T}^{\pi/T} N_k(V_0, \omega) d \omega \right]^2
\right>
\end{equation}
\begin{equation}
= 4 V_0^2 \left(\frac{\eta}{e R \epsilon^2}\right)^2
\left<
\frac{T^2}{4 \pi^2} 
\left[ \int_{-\pi/T}^{\pi/T} \int_{-\pi/T}^{\pi/T} N_k(V_0, \omega)  N^*_k(V_0, \omega') d \omega d \omega' \right]
\right>
\end{equation}
Integration in $\omega$ is limited below $\pi/T$.
We can split the case $\omega = \omega'$ from $\omega \neq \omega'$ to obtain
\begin{equation}
{\sigma_E^2} = 4 V_0^2 \left(\frac{\eta}{e R \epsilon^2}\right)^2\left[
\left<\frac{T}{2 \pi} \int_{-\pi/T}^{\pi/T} \left| N_k(V_0, \omega) \right|^2  d \omega \right>+ \left<\frac{T}{2 \pi} \int_{-\pi/T}^{\pi/T} \int_{\omega' \neq \omega}^{}  N_k(V_0, \omega)  N^*_k(V_0, \omega') d \omega d \omega' \right>\right]
\end{equation}
Noise in uncorrelated, therefore the complex phase of $N_k(V_0, \omega) N^*_k(V_0, \omega')$ is random for $\omega' \neq \omega$, and the second term averages to zero, and we are left with
\begin{equation}
{\sigma_E^2} = 4 V_0^2 \left(\frac{\eta}{e R \epsilon^2}\right)^2
\left<
 \frac{T}{\pi} \left[ \int_{0}^{\pi/T}\left| N_k(V_0, \omega) \right|^2  d \omega \right]
\right>
\end{equation}
\begin{equation}
 = 4 V_0^2 \left(\frac{\eta}{e R \epsilon^2}\right)^2
\frac{T}{\pi}  \int_{0}^{\pi/T}\left| N(V_0, \omega) \right|^2  d \omega
\end{equation}
This is the noise spectrum integrated up to a frequency $\omega = \pi/T$, or $f = 1/(2T)$.
It can be expressed in terms of noise RMS limited to a bandwidth $BW = 1/(2T)$
\begin{equation}
{\sigma_E^2} = 4 V_0^2 \left(\frac{\eta}{e R \epsilon^2}\right)^2
\frac{2 T^2}{\pi} N^2_{RMS}(V_0) |_{BW=1/(2T)}
\end{equation}
Alternatively, assuming a white noise spectral density $e_W$, it can be expressed as
\begin{equation}
{\sigma_E^2} = 4 V_0^2 \left(\frac{\eta}{e R \epsilon^2}\right)^2
\frac{T}{\pi} e_W^2 (V_0) 
\end{equation}
which, by using equation \ref{eq:appenergy1}, can be rewritten as
\begin{equation}
{\sigma_E^2} \simeq \frac{4}{\pi} \frac{\eta}{e R \epsilon^2} e_W^2 (V_0) \overline{E}
\end{equation}
That is the expression for the fluctuation of the injected energy used in the paper, equation \ref{sigmaE2}.

\end{document}